\documentclass[onecolumn]{openjournal}
\usepackage{amsmath}
\usepackage{enumitem}
\usepackage{multirow}
\usepackage{subcaption}
\usepackage[hidelinks]{hyperref}
\usepackage{needspace}
\usepackage{placeins}
\let\OJApOriginalSubsection\subsection
\renewcommand{\subsection}{\Needspace{4\baselineskip}\OJApOriginalSubsection}
\BeforeBegin{subfigure}{\unskip}
\AfterEnd{subfigure}{\ignorespaces}
\bibpunct{(}{)}{;}{a}{}{,}
\expandafter\let\csname longtable*\endcsname\relax
\expandafter\let\csname endlongtable*\endcsname\relax

\makeatletter
\let\OJApOriginalLastPageLabel\lastpage@putlabel
\def\lastpage@putlabel{%
  \global\let\lastpage@putlabel\relax
  \OJApOriginalLastPageLabel}
\makeatother

\shortauthors{Long et al. }
\shorttitle{Cross-database atomic energy-level data mapping}

\begin{document}

\title{Automated Cross-database Mapping of Atomic Energy-Level Data for Astrophysical Plasma Models
}

\author{Qirui Long\altaffilmark{1,*}}
\author{Junjie Mao\altaffilmark{1,*}}
\author{Yihan Tao\altaffilmark{3}}
\author{Xuan Yang\altaffilmark{2,3}}
\author{Dongwei Fan\altaffilmark{3}}
\affil{\altaffilmark{1}Department of Astronomy, Tsinghua University,
             Beijing 100084, China\\
       \altaffilmark{2}School of Astronomy and Space Science,
             University of Chinese Academy of Sciences,
             Beijing 100049, China\\
       \altaffilmark{3}National Astronomical Observatories, Chinese Academy of Sciences,
             Beijing 100101, China}
\altaffiltext{*}{longqr24@mails.tsinghua.edu.cn;
      jmao@mail.tsinghua.edu.cn}

\begin{abstract}
Accurate atomic data are essential for interpreting high-resolution X-ray spectra of astrophysical plasmas. However, different plasma models are based on different atomic databases, which have different energy level data. In this work, we develop an automated cross-database framework for matching atomic energy-level data across AtomDB v3.1.3, CHIANTI v11.0.2, and SPEX v3.08.01. For each pair of databases, we construct similarities from three complementary components: the electronic configuration, the term quantum numbers and parity ($2S+1,~L,~J,~\pi$), and the energy value. The configuration similarity is calculated with a BiLSTM-based model, the similarity of \(2S+1,~L,~J,~\pi\) is calculated from a rule-based comparison of ($2S+1,~L,~J$) and the configuration-derived parity ($\pi$), and the energy similarity is calculated from a three-component energy vector describing the absolute energy, the energy group, and the relative position within the band. These components are combined into a total similarity matrix. A threshold-constrained Hungarian algorithm is then used to obtain a partial matching map, allowing levels without reasonable counterparts to remain unmatched. We apply the method to H-like, He-like, Li-like, Be-like, and Ne-like isoelectronic sequences. The results show that acceptable threshold ranges can be found for these five sequences under six rule-based validation criteria, including exact consistency in the standardized parity label. The resulting maps provide reproducible cross-database energy-level correspondences and form a basis for future comparisons of wavelengths, transition probabilities, collision strengths, and plasma emission models.
\end{abstract}
\keywords{atomic data -- atomic processes -- methods: data analysis -- plasmas -- X-rays: general}
\maketitle

\section{Introduction}           
\label{sect:intro}

In astronomy, astrophysical plasmas refer to high-temperature ionized matter in the universe. They are widely found in stellar coronae, supernova remnants, galaxies, galaxy clusters, and the warm-hot intergalactic medium distributed along the cosmic web \citep{Kaastra2017,Rybicki1979,Kaastra2008}. To study these systems, researchers often use the X-ray emission and absorption of plasmas. Emission lines from different ions in different ionization states are very important tools to constrain plasma properties, such as temperature, density, elemental abundance, velocity structure, and so on \citep{Kaastra2008,Raymond2005}. Plasma models used to fit these spectral lines are built using atomic databases. And the most basic atomic data in databases is the energy-level data.

High-resolution X-ray spectroscopy has progressed from dispersive grating instruments on Chandra and XMM-Newton, such as HETG and RGS, to non-dispersive microcalorimeter instruments such as SXS in Hitomi and Resolve in XRISM \citep{Canizares2005,denHerder2001,Kelley2016SXS,Tashiro2022,Ishisaki2025Resolve}. Future missions such as HUBS and NewAthena are expected to have a further improvement in high-resolution and high-throughput X-ray spectroscopy \citep{Cui2020HUBS,Bregman2023HUBS,Cruise2025}. These facilities can resolve increasingly detailed line structures, which shows stricter requirements on plasma models and the basic atomic data \citep{Hitomi2018}.
Previous high-resolution studies have shown that different versions of atomic databases can lead to noticeable differences in fitted parameters of plasma models, including elemental abundances \citep{Hitomi2018,Plucinsky2025}. 
Similar atomic data and model code dependent effects have also been reported for intracluster medium abundance measurements, Fe-L line emissivities and atomic data constraints, and charge exchange spectra \citep{Mernier2020,Chakraborty2024,Gu2019FeL,Gu2020FeLII,Gu2022CX,Gu2022FeL}.
These results demonstrate that atomic data differences can directly affect spectral fitting results and, consequently, the physical interpretation of astrophysical plasmas.

Most commonly used astrophysical plasma models are built upon their own atomic databases. Astrophysical Plasma Emission Code (APEC) is based on the Astrophysical Plasma Emission Database (APED) in AtomDB \citep{Smith2001,Foster2012,Foster2020PyAtomDB}, the CHIANTI code is based on the CHIANTI atomic database \citep{Dere1997,DelZanna2015,Young2020,DelZanna2021,Dere2023,Dufresne2024}, and SPEX relies on the atomic database in the SPEX Atomic Code \& Tables (SPEXACT)\citep{Kaastra1996,dePlaa2020,SPEX30801,SPEX30803}. Other evaluated or public atomic data resources, such as the NIST Atomic Spectra Database and OPEN-ADAS, together with systematic atomic-data calculations for astrophysical plasma diagnostics, also provide important reference data for atomic energy levels, wavelengths, transition probabilities, and electron-impact excitation data \citep{NISTASD2024,Ralchenko2005NISTASD,Summers2004ADAS,Mao2020a,Mao2020b,Mao2021b,Mao2022}. Atomic structure data, including energy levels and transition data, form the basis for calculating atomic process data, identifying spectral lines, and constructing plasma models; data for various atomic processes, such as dielectronic recombination (DR), radiative recombination (RR), ionization, and collisional excitation, are also mapped based on the underlying structure data \citep{Kaastra2008,Foster2012}. Therefore, to understand the differences between different plasma models, it is necessary to trace them back to the differences between atomic structure data in different databases.

\begin{figure}[hbt!]
    \centering

    \begin{subfigure}[t]{0.30\linewidth}
        \centering
        \includegraphics[width=\linewidth]{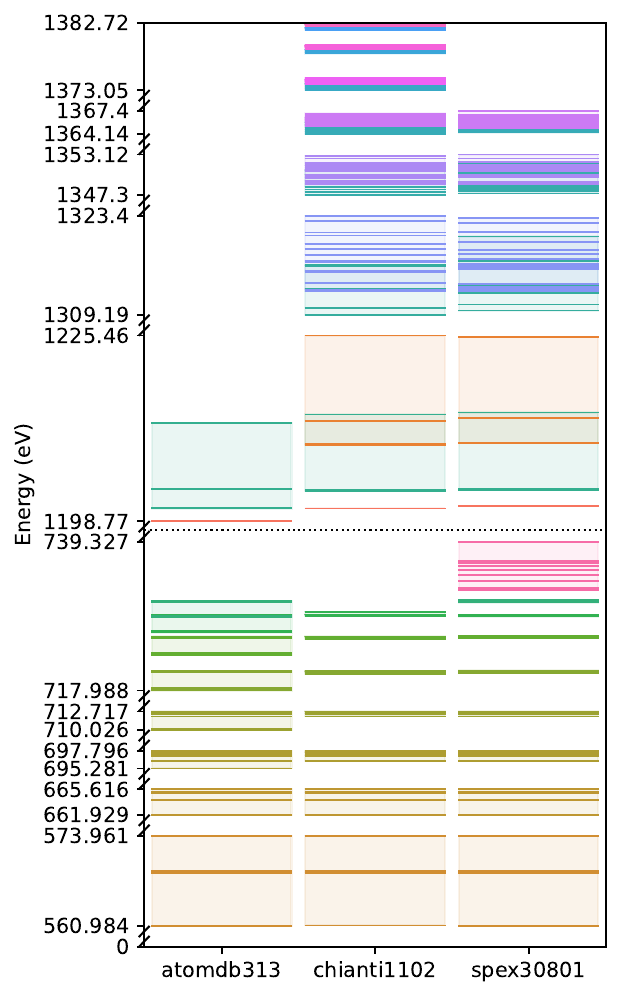}
        \caption{Full distribution.}
        \label{fig:evsdb_full_no_labels}
    \end{subfigure}
    \hfill
    \begin{subfigure}[t]{0.342\linewidth}
        \centering
        \includegraphics[width=\linewidth]{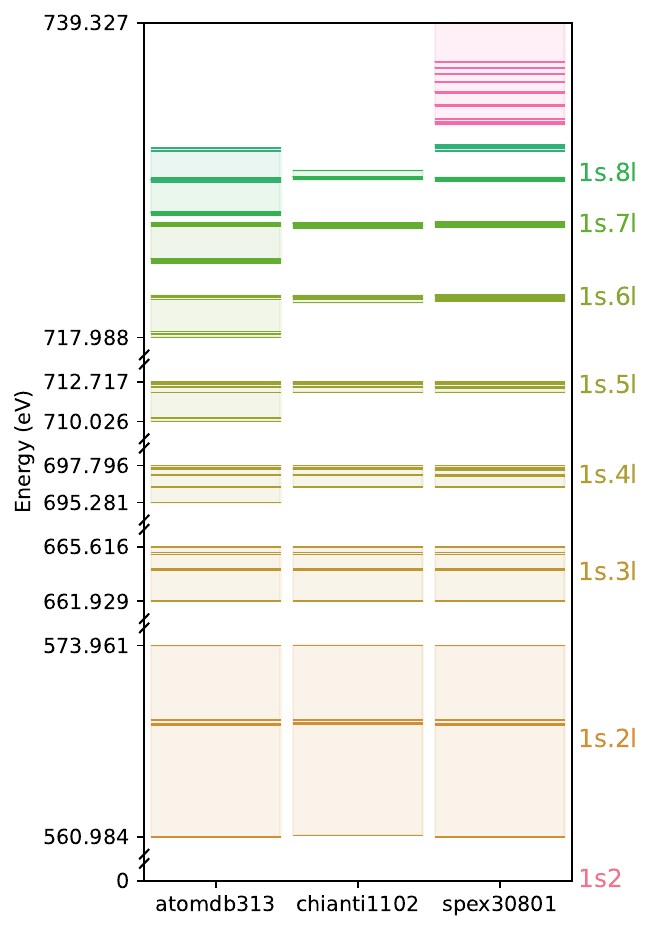}
        \caption{Low-energy distribution.}
        \label{fig:evsdb_low_with_labels}
    \end{subfigure}
    \hfill
    \begin{subfigure}[t]{0.342\linewidth}
        \centering
        \includegraphics[width=\linewidth]{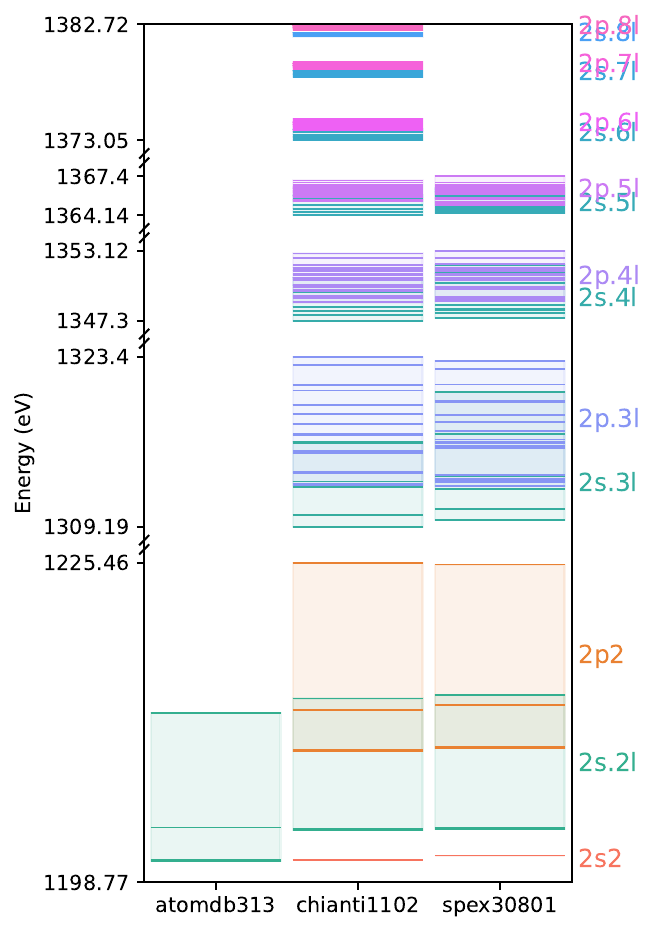}
        \caption{High-energy distribution.}
        \label{fig:evsdb_high_with_labels}
    \end{subfigure}

    \caption{The energy-level distributions of O \textsc{vii} ion in different databases, including AtomDB v3.1.3, CHIANTI v11.0.2, and SPEX v3.08.01. Each solid line represents one energy level, and different colors indicate different level categories classified according to the principal quantum number $n$ in configuration. Panel (a) shows the full energy-level distribution, while panels (b) and (c) show the low-energy and high-energy regions with configurations, respectively.}
    \label{fig:evsdb_three_panels}
\end{figure}

Among atomic structure data, energy-level correspondence is the prerequisite for subsequent comparisons of transitions and other atomic process data. However, cross-database level matching is non-trivial. Different databases may contain different numbers of levels, cover different energy ranges, or have different indices, energies, configurations, and quantum numbers for the same physical level. Conversely, multiple levels may share very similar configurations and quantum numbers, especially in dense or mixed level regions. As illustrated in Fig.~\ref{fig:evsdb_three_panels}, such differences make it difficult to match cross-database energy-level data based only on level indices, nearest energies, configurations, and quantum numbers.

Recurrent neural network based models have been applied to astronomical time-domain classification, for example in the classification of unevenly sampled variable-star light curves \citep{Naul2018}. In astronomical cross-identification, the matching task can be formulated as a global assignment problem. \citet{BudavariBasu2016} used the Hungarian algorithm for Bayesian two-catalog cross-identification, while \citet{Shi2019} developed a closely related integer linear programming assignment framework for multiple catalogs.
These studies show that machine learning methods and Hungarian algorithm based assignment are feasible tools to solve astronomical problems.

In this work, we propose an automated workflow to match cross-database energy-level data. For each pair of databases, we compute similarities by three components: the electronic configuration, the term quantum numbers and parity ($2S+1,~L,~J,~\pi$), and the energy value. The configuration similarity is obtained with a Bidirectional Long Short-Term Memory (BiLSTM) model \citep{Hochreiter1997,Schuster1997}, the similarity of \(2S+1,~L,~J,~\pi\) is calculated from a rule-based comparison of ($2S+1,~L,~J$) and the parity ($\pi$), and the similarity of energy value $E$ is calculated from an energy vector describing the absolute energy, the energy group, and the relative position within the band. Then we obtain the final one-to-one matching map with a threshold-constrained Hungarian algorithm \citep{Kuhn1955,Munkres1957}. We apply the method to AtomDB v3.1.3, CHIANTI v11.0.2, and SPEX v3.08.01 for the H-like, He-like, Li-like, Be-like, and Ne-like isoelectronic sequences. We do not aim to determine which database is ``correct''. We provide reproducible energy-level correspondences, as a basis for future comparisons of atomic data and plasma models. For more complex isoelectronic sequences, future work will explore machine learning based level matching that uses not only energy-level data but also transition data to construct more reliable cross-database level maps.


\section{Atomic Data}

\subsection{Data Sources}

In this work we use atomic energy-level data from three widely used astrophysical atomic databases: AtomDB v3.1.3, CHIANTI v11.0.2, and SPEX v3.08.01. AtomDB data were obtained from the APED, which provides atomic data for the APEC \citep{Smith2001,Foster2012,Foster2020PyAtomDB}.\footnote{\url{https://www.atomdb.org/}} CHIANTI data were taken from the stand-alone CHIANTI atomic database v11.0.2 \citep{Dere1997,DelZanna2015,Young2020,DelZanna2021,Dere2023,Dufresne2024}.\footnote{\url{https://www.chiantidatabase.org/chianti_download.html}} 
SPEX level data were extracted from the atomic database distributed with SPEXACT (SPEX v3.08.01) \citep{Kaastra1996,dePlaa2020,SPEX30801}.\footnote{\url{https://zenodo.org/records/12771915}} 

\subsection{Standardized Format and Data Cleaning}

The original databases use different file formats and column definitions. 
AtomDB and SPEX provide the atomic energy-level data in FITS files \citep{Wells1981,Pence2010}, while CHIANTI provides ASCII database files. 
We converted all level tables into a common format before matching. The standardized columns are listed in Table~\ref{tab:level_format}. 
For CHIANTI v11.0.2, we use the observed energy when a valid value is available and the theoretical energy otherwise. The only exception is level 81 of Na \textsc{ix}, whose observed energy appears to be erroneous, differing from the corresponding AtomDB value by approximately 90\%. Therefore, we use the theoretical energy instead.

\begin{table}[!hbt]
    \centering
    \begin{tabular}{ccccccccc}
        \hline
        idx & cfg & $2S+1$ & $L$ & $J$ & $\pi$ &  pcm & ryd & eV\\
        \hline
        1 & 1s2 & 1 & 0 & 0 & 0 & 0 & 0 & 0 \\
        2 & 1s.2s & 3 & 0 & 1 & 0 & 4524640.0 & 41.231552 & 560.9839 \\
        ... & ... & ... & ... & ... & ... & ... & ... \\
        116 & 1s.8i & 1 & 6 & 6 & 0 & 5859404.5 & 53.39482 & 726.4736 \\
        ... & ... & ... & ... & ... & ... & ... & ... \\
        204 & 2s.2p & 1 & 1 & 1 & 1 & 9782607.0 & 89.145676 & 1212.8887 \\
        \hline
    \end{tabular}
    \caption{Example of the standardized energy-level data of O \textsc{vii} from AtomDB v3.1.3. \texttt{idx} is the level index in the database; \texttt{cfg} is the electronic configuration; $2S+1$, $L$, and $J$ are quantum numbers, respectively the spin multiplicity, total orbital angular momentum, and total angular momentum; and $\pi$ is the parity. The parity is calculated from the configuration using Eq.~\ref{eq:parity}. The energy value is stored in three units: cm$^{-1}$ (pcm), Rydberg (ryd), and eV. The $L$ symbols were converted to numerical values, with S, P, D, F, $\ldots$ corresponding to $L=0,1,2,3,\ldots$.}
    \label{tab:level_format}
\end{table}

The converted data were stored in HDF5 format, which is a hierarchical binary format that supports compression, metadata, and efficient random access \citep{Folk2011,HDFGroup2025}. 
This format is more compact and structured than ASCII files, and more convenient than separate FITS tables for storing energy-level data from several databases for the same ion. 
Each ion is saved as one HDF5 file, with separate datasets such as \texttt{lev\_atomdb313}, \texttt{lev\_chianti1102}, and \texttt{lev\_spex30801}.
 
Before matching, trailing dot padding and any whitespace following it were removed from configuration strings. For example, `1s....` was standardized to `1s`. Internal orbital separators and parent-state labels were preserved. For AtomDB, negative values of $2S+1$ or $L$ were excluded from the standardized energy-level data because they do not provide usable term information for the similarity calculation. The parity $\pi$ is calculated from the electronic configuration as
\begin{equation}
\pi = \left(\sum_k n_k l_k\right) \bmod 2 , \label{eq:parity}
\end{equation}
where $n_k$ is the occupation number of subshell $k$, and $l_k$ is its orbital angular momentum quantum number. Equation~\ref{eq:parity} is applied to ordinary bound-level configurations, for which $\pi=0$ represents even parity and $\pi=1$ represents odd parity. For entries containing \texttt{limit}, we assign the reserved standardized label $\pi=2$. This value is not a physical parity; it distinguishes limit entries from ordinary even- and odd-parity levels. If the parity of a non-limit configuration cannot be determined, $\pi=-1$ is assigned. Figure 2 counts the retained levels in standardized energy-level data used as matching inputs.

\begin{figure}[hbt!]
    \centering
    \includegraphics[width=0.85\linewidth]{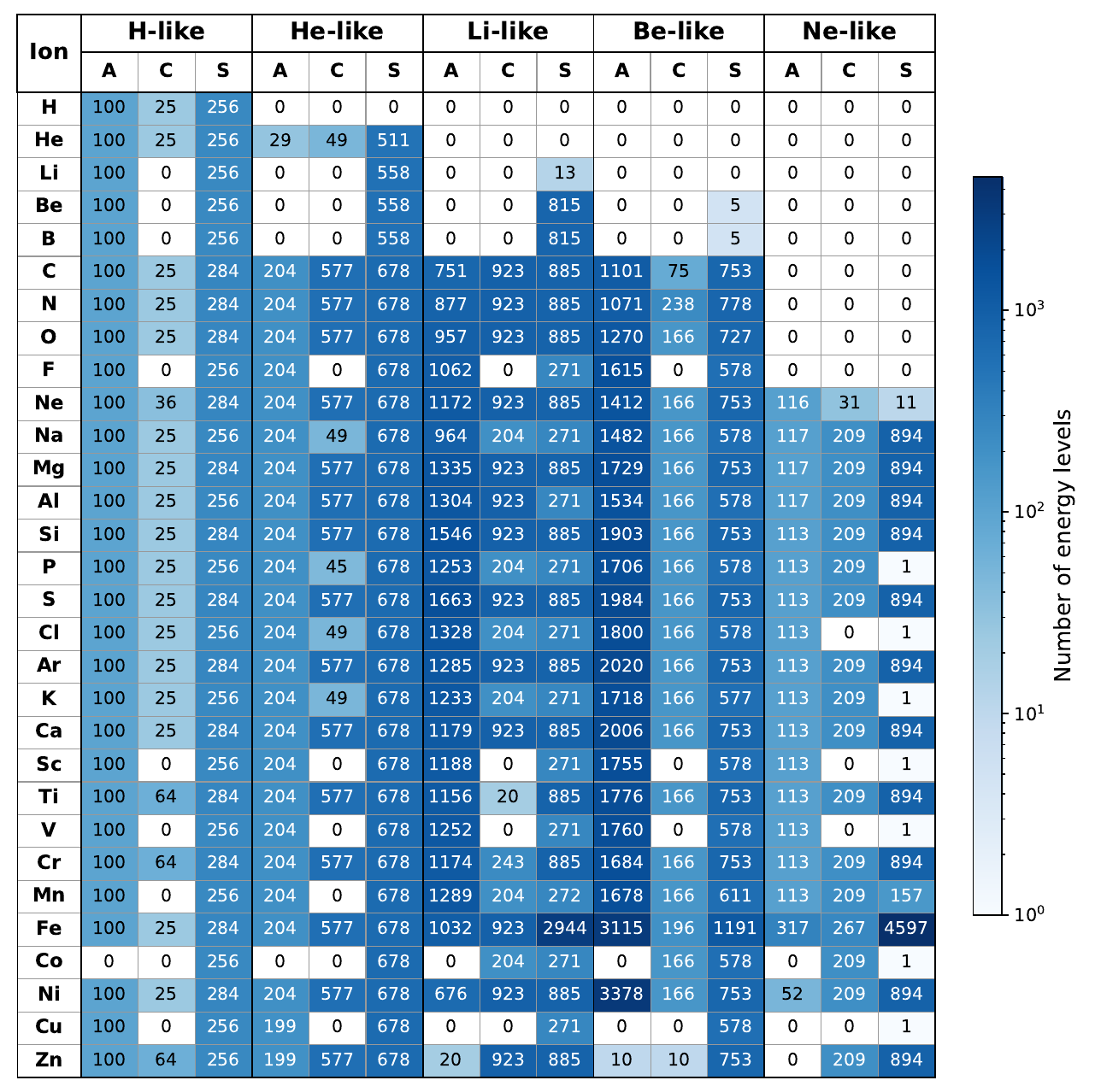}
    \caption{Number of standardized energy-level data used as matching inputs in this work. The “A,” “C,” and “S” in the second row represent AtomDB v3.1.3, CHIANTI v11.0.2, and SPEX v3.08.01, respectively.}
    \label{fig:level_number}
\end{figure}

\section{Method}
\label{sec:method}

The goal of this work is to obtain the matching map between the energy-level data in different atomic databases. We decompose each energy-level data entry into three components: the electronic configuration, the term quantum numbers and parity $(2S+1,~L,~J,~\pi)$, and the energy value. The configuration similarity is described in Sect.~\ref{sec:cfg_similarity}. We train a configuration model based on a Bidirectional Long Short-Term Memory (BiLSTM) encoder \citep{Hochreiter1997,Schuster1997} to measure configuration similarity across databases. The similarity of \(2S+1,~L,~J,~\pi\) is described in Sect.~\ref{sec:term_parity_similarity}. It is calculated from a rule-based comparison of $2S+1,~L,~J$ and the parity $\pi$. The energy similarity is described in Sect.~\ref{sec:energy_similarity}. We represent energy value $E$ as a feature vector with three components: the nonlinearly transformed absolute energy value, the energy group to which $E$ belongs, and the relative position within that group. The energy similarity is then calculated from this energy feature vector.

The three similarity components are combined into a total similarity matrix, as described in Sect.~\ref{sec:total_similarity_hungarian}. Since not every level in one database has a physical counterpart in another database, we introduce a threshold to avoid forced low similarity matches. Then we used the threshold-constrained Hungarian algorithm \citep{Kuhn1955,Munkres1957} to obtain the globally optimal one-to-one matching map.

\subsection{Configuration Similarity}
\label{sec:cfg_similarity}

The electronic configuration describes how the electrons of an ion occupy atomic orbitals \citep{Cowan1981}. For example, $\mathrm{1s.2p}$ means that one electron occupies the 1s orbital, with $n=1$ and $l=0$, and another electron occupies the 2p orbital, with $n=2$ and $l=1$. Energy levels with the same or physically similar configurations should have a high configuration similarity. However, configuration formats are not uniform among databases, as shown in Table~\ref{tab:cfg_examples}. AtomDB usually gives a full configuration, while CHIANTI and SPEX may omit closed-shell cores, for example $\mathrm{2p}$ in Table~\ref{tab:cfg_examples}. In addition, CHIANTI and SPEX may contain parent-state labels in parentheses, shown in Table~\ref{tab:cfg_examples}. Some configuration labels of SPEX also contain the marker \texttt{limit}, which denotes an ionization limit associated with a specific level of the next higher ionization stage rather than a normal bound-level configuration. Such data with the marker \texttt{limit} are assigned the parity label $\pi$=2.
These format differences make exact configuration string matching unsuitable for cross-database comparison. Furthermore, in addition to these three databases, we want to develop a method that can be used in other databases that have corresponding configurations.

\begin{table}[h]
\centering
\caption{Configuration notations in different databases. The contents of the three columns ‘AtomDB v3.1.3’, ‘CHIANTI v11.0.2’ and ‘SPEX v3.08.01’ are the standardized configuration notations. The marker \texttt{limit}, as in 2p.limit, denotes an ionization limit. “-” indicates that there is no physically corresponding configuration in the database}
\label{tab:cfg_examples}
\begin{tabular}{cccc}
\hline
Sequence & AtomDB v3.1.3 & CHIANTI v11.0.2 & SPEX v3.08.01 \\
\hline
Li-like & $\mathrm{1s2.2p}$ & $\mathrm{2p}$ & $\mathrm{2p}$ \\
Li-like & $\mathrm{1s.2p.6g}$ & $\mathrm{1s.2p(3P).6g}$ & $\mathrm{1s.2p.(3P).6g}$ \\
Be-like & $\mathrm{1s.2s.2p2}$ & $\mathrm{1s.2s.2p2}$ & $\mathrm{1s.(2s.2p2.(2D))}$ \\
Be-like & - & - & 2p.limit \\
Ne-like & - & - & $\mathrm{2s2.2p4.(3P).3p2.(3P)}$ \\
\hline
\end{tabular}
\end{table}

To obtain a configuration similarity that can be applied to different database pairs, we trained a configuration model based on a Bidirectional Long Short-Term Memory (BiLSTM) encoder. LSTM is an improved recurrent neural network (RNN) architecture \citep{Hochreiter1997,Schuster1997}. A BiLSTM reads a sequence in both forward and backward directions, so the representation of each character contains context from both sides of the string \citep{Hochreiter1997,Schuster1997}. BiLSTM is useful for this work because configuration similarity depends not only on individual orbital notations, but also on their order and database-dependent notation, such as omitting closed shells and parent states. This model treats each configuration as a character sequence. For level matching, the model receives the standardized configuration strings described in Sect. 2.2. In modeling, the characters of configuration strings are first encoded as fixed-length sequences of character indices, and then passed through an embedding layer and a bidirectional LSTM. Then the output sequence is averaged and projected into a configuration vector. So for a configuration $c$, the model gives an embedding vector
\begin{equation}
    \mathbf{x}_{\rm cfg}(c) = f_{\rm cfg}(c).
\end{equation}
The configuration similarity between two levels $i$ and $j$ from two different databases is then calculated using the cosine similarity \citep{Salton1983} 
\begin{equation}
    S_{\rm cfg}(i,j) =
    \frac{
    \mathbf{x}_{\rm cfg}(c_i) \cdot \mathbf{x}_{\rm cfg}(c_j)
    }{
    \left|\mathbf{x}_{\rm cfg}(c_i)\right|
    \left|\mathbf{x}_{\rm cfg}(c_j)\right|
    } .\label{eq:cos_cfg}
\end{equation}

The training pairs were generated from real configurations extracted from the standardized energy-level data. For each isoelectronic sequence, we generated positive pairs with different similarity levels. We aimed to assign high similarities to configurations with the same physical meaning across different databases and low similarities to configurations with different physical meanings. The training examples were therefore divided into five similarity levels. There are four positive relation types: hard, shortening, medium, and soft, together with negative examples sampled dynamically during training. The pair types are summarized in Table~\ref{tab:cfg_pair_types}. Hard pairs represent identical or format-equivalent configurations and are assigned the highest weight. Shortening pairs omit closed-shell prefixes from the configuration. Medium pairs are generated by partially or completely removing parent-state labels. Soft pairs describe configurations that differ only in the last orbital angular-momentum symbol ($|\Delta l|=1$) within the same structural category.

The numbers of training pairs for the five isoelectronic sequences are listed in Table~\ref{tab:pair_statistics}. 
We extract configurations of five isoelectronic sequences from all databases. By performing format transformations, omitting closed-shell prefixes, removing parent-state labels, and changing the angular momentum symbol of the last orbital, we construct four types of positive training pairs and assign different weights to them.
All training pairs are constructed from database configurations and their allowed variants. We transform complex configuration formats into simpler ones; for example, we pair 1s.2p.(3P).6g with 1s.2p.6g. We also construct pairs by omitting closed-shell prefixes, such as 2p-1s2.2p, and changing $l$, such as 3p-3d. We retain both forward and reverse training pairs, such as 2p-1s2.2p and 1s2.2p-2p, and remove duplicate pairs.
During training, we dynamically select negative examples from the configurations included in these training pairs. 
We give priority to configurations whose last-orbital principal quantum number differs from that of the anchor by at least two ($|\Delta n|\geq 2$), and we try to avoid selecting configurations that differ only in parent-state or closed-shell notation. These negative examples are not included in Table~\ref{tab:pair_statistics}.

\begin{table}[hbt!]
\centering
\caption{Training pair types used for the configuration model.}
\label{tab:cfg_pair_types}
\begin{tabular}{lccc}
\hline
Pair type & Weight & Sample configuration pair  & Meaning \\
\hline
Hard & 1.00 & 1s.2p.(3P).6g - 1s.2p(3P).6g & Identical or format-equivalent configurations \\
Shortening & 0.95 & 1s2.2p - 2p & Closed-core omission \\
Medium & 0.80 & 1s.2p.(3P).6g - 1s.2p.6g & Parent-state labels partly or fully removed \\
Soft & 0.10 & 1s2.3p - 1s2.3d & Same structure class, final orbital $|\Delta l|=1$ \\
Negative & -- & 1s2.3p - 1s2.8d & Explicitly sampled dissimilar configurations \\
\hline
\end{tabular}
\end{table}

\begin{table}[hbt!]
\centering
\caption{Statistics of the positive training pairs for different isoelectronic sequences. Negative pairs were sampled dynamically during model training and are not included in this table.}
\label{tab:pair_statistics}
\begin{tabular}{lrrrrr}
\hline
Sequence & All & Hard & Shortening & Medium & Soft \\
\hline
H-like  & 355  & 137  & 0    & 0    & 218 \\
He-like & 479  & 187  & 0    & 0    & 292 \\
Li-like & 3529 & 1701 & 68   & 1036 & 724 \\
Be-like & 1396 & 464  & 236  & 8    & 688 \\
Ne-like & 5329 & 1993 & 1178 & 1480 & 678 \\
\hline
\end{tabular}
\end{table}

For each of the five isoelectronic sequences, we generated an independent training set and trained a separate configuration model. The similarities calculated by these models are shown in the diagnostic plots below.
Our model first pads or truncates the configuration strings to a length of 30 characters, then extracts features using character embeddings and a bidirectional LSTM, ultimately yielding a 128-dimensional configuration vector. We used the Adam optimizer with a learning rate of \(10^{-3}\), a batch size of 256, and a total of 100 training epochs during training. We combine a weighted contrastive loss and a triplet loss into a composite objective function, with coefficients of 1 and 0.5, respectively. This loss function encouraged each anchor configuration to have a higher cosine similarity to its positive counterpart than to negative examples.
After training, the corresponding configuration model can be used to compute $S_{\rm cfg}$ for any energy-level pair from two given databases in the same isoelectronic sequence.

\begin{figure}[hbt!]
    \centering
    \includegraphics[width=0.7\linewidth]{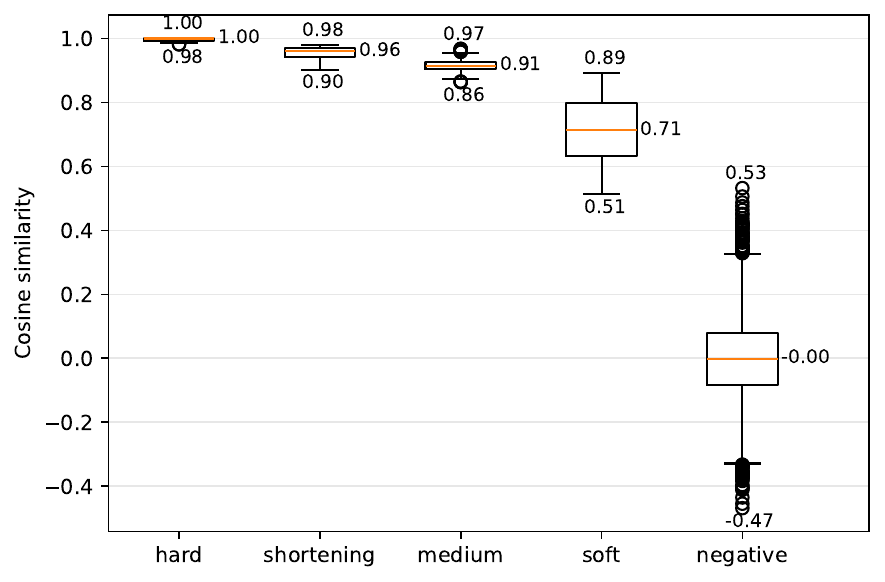}
    \caption{This boxplot shows the cosine-similarity distributions of hard, shortening, medium, soft, and negative configuration pairs in Li-like ions. Positive pairs are taken from the configuration training pairs, and negative pairs are sampled from the same configuration pool. The similarity values are computed from the trained configuration model. In each box, the orange line marks the median, and the lower and upper box edges indicate the first and third quartiles, respectively. The numerical labels indicate the minimum, median, and maximum similarity values. Open circles represent outliers beyond the whiskers. }
    \label{fig:cfg_cosine_boxplot}
\end{figure}

We used two diagnostic plots to check the behavior of the trained model. Fig.~\ref{fig:cfg_cosine_boxplot} shows the distribution of cosine similarities for different relation types in Li-like ions. The learned similarities generally follow the designed hierarchy: hard pairs have the highest similarities, close to \(1\), followed by shortening pairs with similarities about $~0.96$, then medium pairs with similarities about $~0.91$, while soft pairs have similarities about $~0.71$, and negative pairs are mostly separated from the positive groups. Small overlaps are expected because the rule-based relation types are not strictly separable from each other. Configurations in some neighboring groups are similar, so the model may learn similar representations for them.

\begin{figure}[hbt!]
    \centering
    \includegraphics[width=0.98\linewidth]{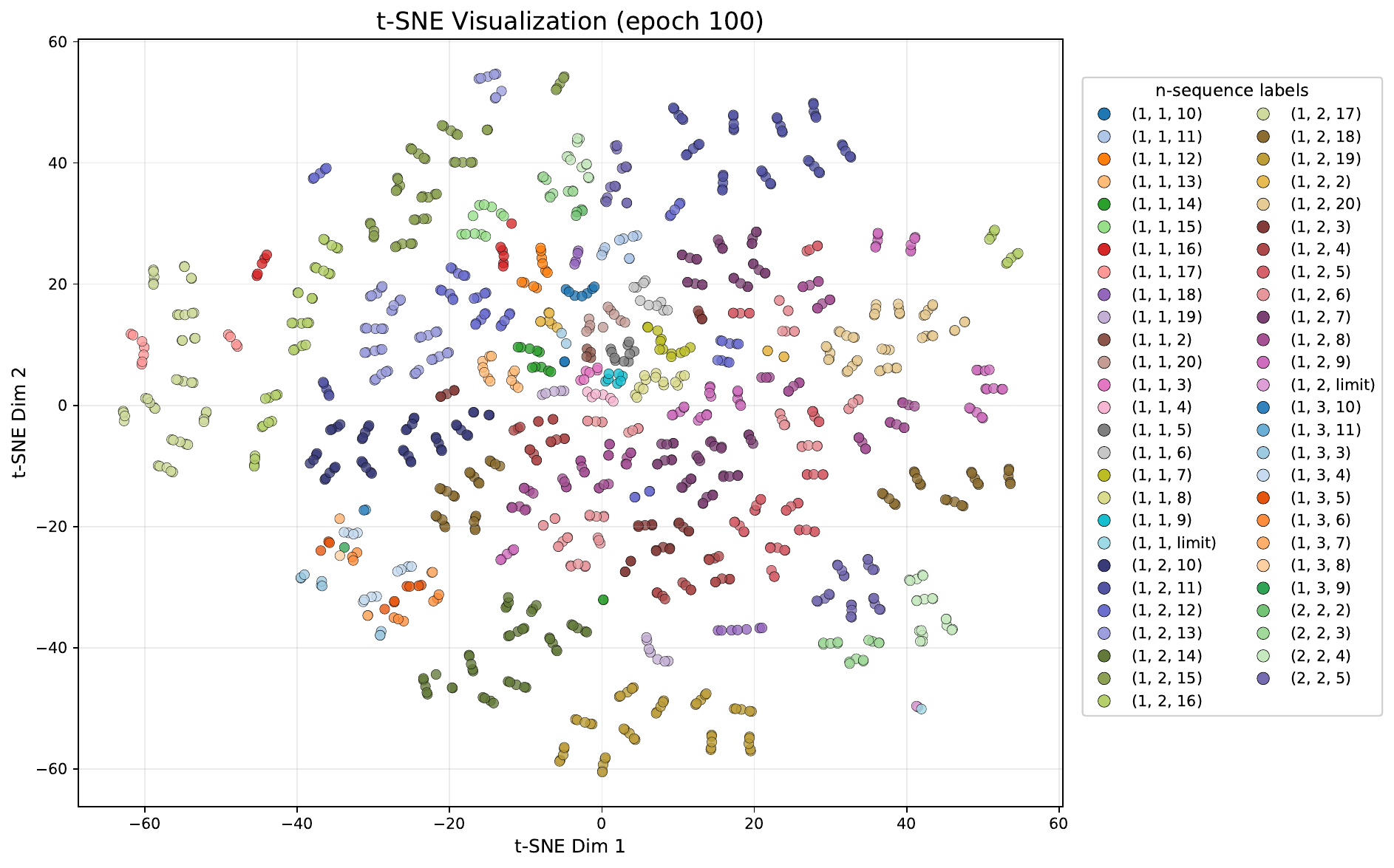}
    \caption{This t-SNE plot visualizes the configuration embeddings learned by the trained configuration model in Li-like ions. Each point represents one unique configuration from the training pairs. During plotting, up to 1200 configurations are randomly sampled, encoded by the configuration model, and projected into two dimensions using t-SNE. The colors indicate different $n$ sequence labels, which are obtained from the principal quantum numbers in each configuration. For example, $(1,~2,~3)$ means $1l2l3l$, and 1s2s3p and 1s2p3p both belong to $1l2l3l$.}
    \label{fig:cfg_tsne}
\end{figure}

Figure~\ref{fig:cfg_tsne} shows a t-distributed stochastic neighbor embedding (t-SNE) projection of the configuration embeddings after training \citep{vanderMaaten2008}. 
Configurations with the same $n$ sequence labels tend to appear close to each other. This suggests that the learned embedding captures similarities in the principal quantum number structure of the configurations. The scatter in the figure is expected because t-SNE is only a two-dimensional projection of the learned high-dimensional embedding space and there are even finer-grained configuration structures in addition to $n$. 

\subsection{Similarity of \texorpdfstring{$2S+1$, $L$, $J$, and $\pi$}{2S+1, L, J, and Parity}}
\label{sec:term_parity_similarity}

The quantities $2S+1$, $L$, and $J$ describe the spin multiplicity, total orbital angular momentum, and total angular momentum of an atomic level, respectively. They provide discrete quantum number constraints for identifying the physical state of the energy level. 
In addition, the parity $\pi$ of an energy level is an important atomic property, which describes whether the wave function of an atomic state remains unchanged or changes sign after spatial inversion. 
After standardization, $2S+1$, $L$, $J$ and parity are numerical quantities. In general, these quantities are expected to be identical in different databases for the same physical energy level. We therefore use a direct pairwise comparison to calculate their similarity.

The similarity based on $2S+1$, $L$, $J$, and parity for a level pair $(i,j)$ is defined by a rule-based score as follow. There are four situations: when all four quantities are identical, it is considered a full match which $S_{SLJP}(i,j)=1.0$; if $J$ and parity $\pi$ are identical, but there is a slight difference in either $M$ or $L$ between the two levels, $S_{SLJP}(i,j)=0.1$; if both small differences occur at the same time, $S_{SLJP}(i,j)=0.01$; all other candidate pairs are assigned $S_{SLJP}(i,j)=0$. 
\begin{equation}
S_{SLJP}(i,j)=
\begin{cases}
1.0, & M_i=M_j,\ L_i=L_j,\ J_i=J_j,\ \pi_i=\pi_j,\\
0.1, & J_i=J_j,\ \pi_i=\pi_j,\ 
\begin{aligned}
&(|M_i-M_j|=2,\ L_i=L_j)
~\mathrm{or}\ (M_i=M_j,\ |L_i-L_j|=1),
\end{aligned}\\
0.01, & J_i=J_j,\ \pi_i=\pi_j,\ |M_i-M_j|=2,\ |L_i-L_j|=1,\\
0, & \text{otherwise}.
\end{cases}
\end{equation}
where $M=2S+1$.We also apply a physical validity check to $J$. A level is considered valid only if
\begin{equation}
|L-S| \leq J \leq L+S .
\end{equation}
A parity pair is considered valid only when both standardized labels belong to $\{0,1,2\}$ and are identical. Thus, two limit levels may be compared through $\pi_i=\pi_j=2$, whereas a limit level cannot match an ordinary even- or odd-parity level. 
If any level in a pair of candidate configurations does not satisfy the $J$ validity condition or has $\pi=-1$ , the corresponding $S_{SLJP}$ value is set to zero. This prevents situations that violate physical laws from affecting the final matching score.

\subsection{Energy Similarity}
\label{sec:energy_similarity}

\begin{figure}[hbt!]
    \centering

    \begin{subfigure}[t]{0.49\linewidth}
        \centering
        \includegraphics[width=\linewidth]{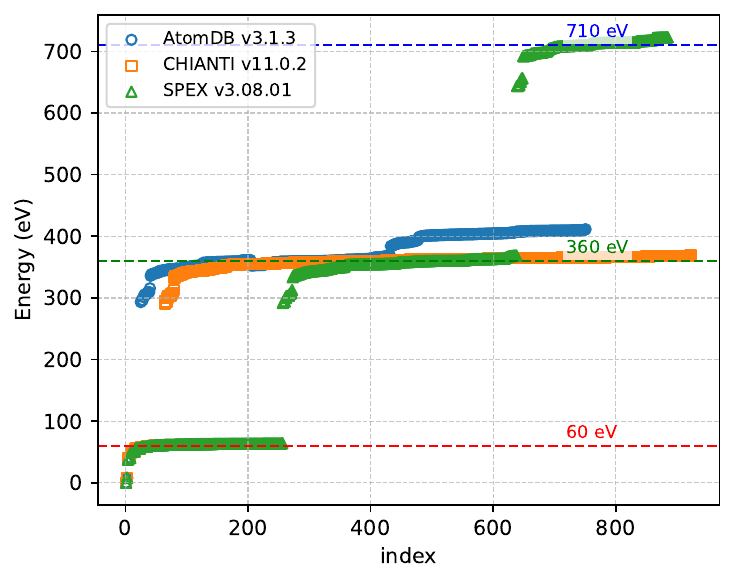}
        \caption{Energy distribution.}
        \label{fig:energy_vs_idx1}
    \end{subfigure}
    \hfill
    \begin{subfigure}[t]{0.49\linewidth}
        \centering
        \includegraphics[width=\linewidth]{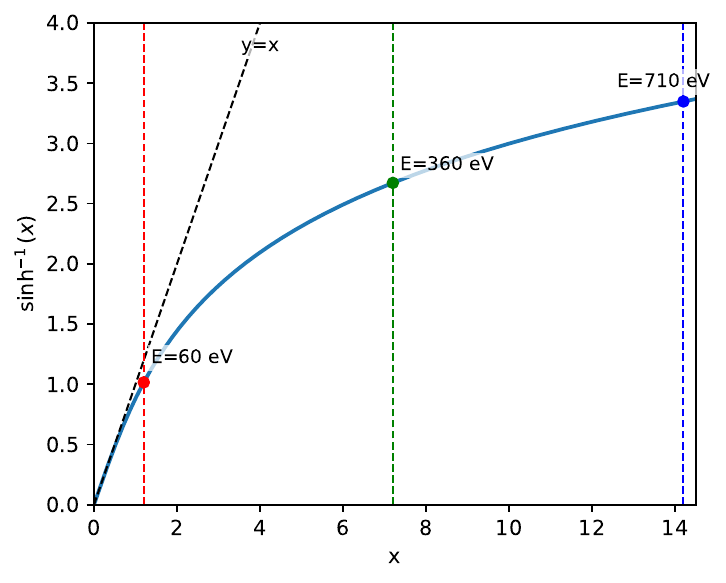}
        \caption{$\sinh^{-1} (E/E_0)$ transformation of $E$.}
        \label{fig/asinh_plot}
    \end{subfigure}

    \caption{The energy value distributions of C \textsc{iv} ion. Panel (a) shows the energy value vs. energy-level index in each database. There are energy groups near \(E=60~\mathrm{eV}\), \(360~\mathrm{eV}\), and \(710~\mathrm{eV}\), which are indicated by dashed lines. The blue open circles, orange open squares, and green open triangles represent the energy-level data from AtomDB v3.1.3, CHIANTI v11.0.2, and SPEX v3.08.01, respectively. Panel (b) shows the transformation of $E$ in the first component. The solid blue line represents $y=\sinh^{-1} (E/E_0)$, where $E_0=50~\text{eV}$, and the dashed black line represents the original $y = x$. The same three energies are marked with dashed lines. }
    \label{fig:energy_vs_idx}
\end{figure}

The energy value $E$ is a very important part of the level description. A direct comparison of one-dimensional $E$ is useful, but it is not sufficient to characterize the similarity of energy value. 
Energy levels are often unevenly distributed in energy value $E$. Some regions are sparse, whereas others are dense, as shown in Fig.~\ref{fig:energy_vs_idx1}. In dense regions, the differences between adjacent energy values may be small, but this does not mean that these differences are insignificant. Small energy differences can help distinguish otherwise similar energy levels in dense regions. To describe $E$ more completely, we first divide the $E$ distribution of each database pair into several energy groups separated by large energy value gaps. Each energy value $E$ is then converted into a feature vector with three components: the transformed absolute energy value, the energy group, and the relative position inside the group.

The first component describes the absolute scale of $E$. To reduce the impact of high energy values at high energy levels, we apply the nonlinear transformation shown in Fig.~\ref{fig/asinh_plot}. We then normalized it, so the first component $a(E)$ is as follows:
\begin{equation}
a(E)=
\frac{
\sinh^{-1}(E/E_0)-\sinh^{-1}(E_\text{min}/E_0)
}{
\sinh^{-1}(E_\text{max}/E_0)-\sinh^{-1}(E_\text{min}/E_0)
},
\end{equation}
where \(E_0=50~\mathrm{eV}\) is a fixed value for all ions and all sequences. Here, \(E_\text{min}\) and \(E_\text{max}\) are respectively the minimum $E$ and the maximum $E$ of the two databases together. The value of \(a(E)\) is clipped to the range \([0,1]\).

The second component describes which energy group $E$ belongs to. For each pair of databases, we first build a common energy background using all $E$ from the two databases. $E$ are sorted, and large gaps are used to divide them into energy groups. The gap threshold is defined empirically as $T_{\rm band} = \max(20~{\rm eV}, 0.05 E_{95})$, where \(E_{95}\) is the 95th percentile of the energy background. If the gap between two adjacent energy values is larger than or equal to \(T_{\rm band}\), a new band is started. For the \(k\)-th band, we denote its lower and upper energy boundaries as \(L_k\) and \(U_k\). If there are \(K\) energy groups, we use a one-hot vector
\begin{equation}
\mathbf{b}(E) = (b_1, b_2, \ldots, b_K),
\end{equation}
where \(b_k=1\) if the level is in the \(k\)-th band, and \(b_k=0\) otherwise. This component helps distinguish levels that are located in different separated energy regions.

The third component describes the relative position of $E$ within its group:
\begin{equation}
r(E)=
\frac{E-L_k}{U_k-L_k},
\end{equation}
for \(E\in[L_k,U_k]\). 
If a band contains only one energy value, we set \(r(E)=0.5\). 
The value of \(r(E)\) is also clipped to the range \([0,1]\).

The final energy feature vector is
\begin{equation}
\mathbf{x}_{E}(E)
=
\left[
w_1 a(E),~
w_2 \mathbf{b}(E),~
w_3 r(E)
\right],
\end{equation}
where the default weights are
$(w_1,w_2,w_3)=(0.4,0.2,0.4)$. These weights are based on empirical findings derived from experiments. The energy similarity between two levels \(i\) and \(j\) from different databases is then calculated with the cosine similarity:
\begin{equation}
S_E(i,j)
=
\frac{
\mathbf{x}_{E}(E_i)\cdot \mathbf{x}_{E}(E_j)
}{
|\mathbf{x}_{E}(E_i)|~|\mathbf{x}_{E}(E_j)|
}.
\end{equation}
This has the same form as Eq.~\ref{eq:cos_cfg}. This construction of the energy feature vector allows the energy similarity to use both global and local information. The transformed absolute energy value preserves the overall energy scale, the group vector describes the energy region, and the relative position helps compare levels inside the same energy group.

\subsection{Total Similarity and Threshold-Constrained Hungarian Matching}
\label{sec:total_similarity_hungarian}

\begin{figure}[hbt!]
    \centering

    \begin{subfigure}[t]{0.49\linewidth}
        \centering
        \includegraphics[width=\linewidth]{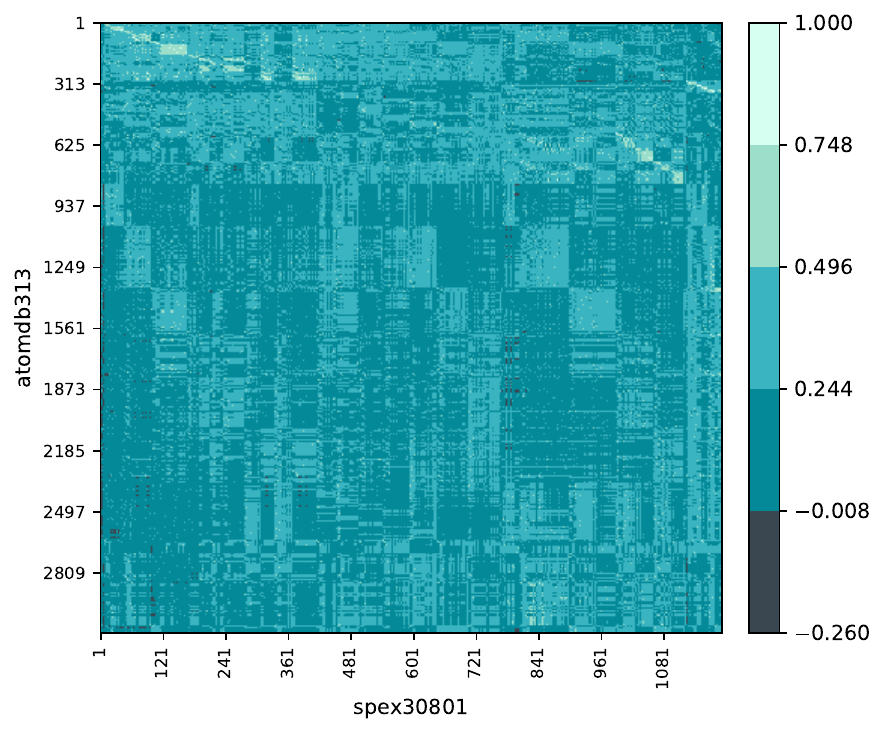}
        \caption{Full similarity matrix.}
        \label{fig:sim_only_be-like_042623_atomdb313_vs_spex30801_db1idx_all-all_db2idx_all-all}
    \end{subfigure}
    \hfill
    \begin{subfigure}[t]{0.49\linewidth}
        \centering
        \includegraphics[width=\linewidth]{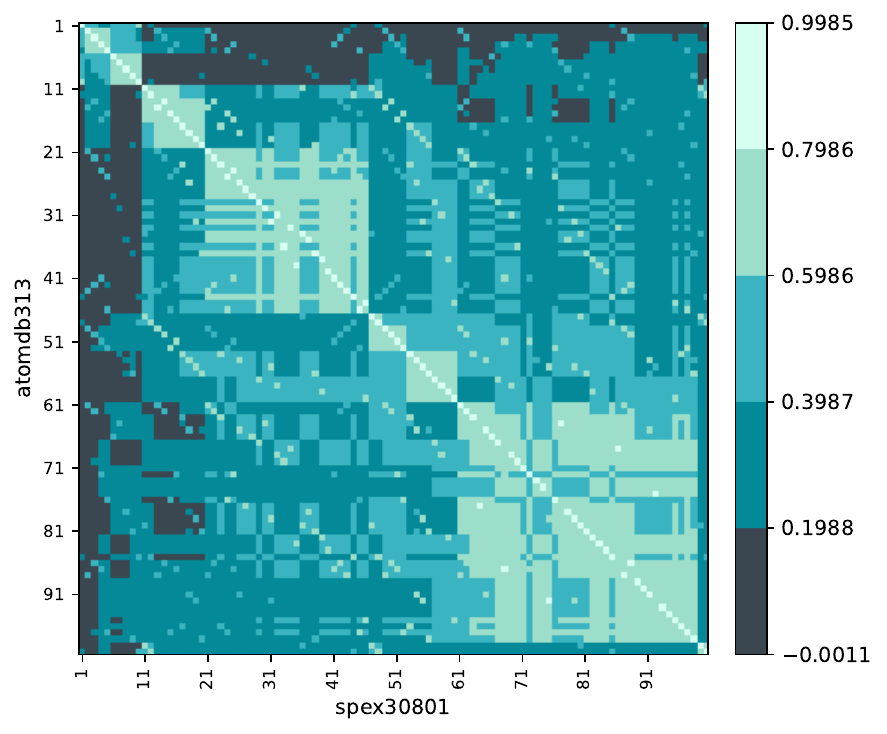}
        \caption{Part of similarity matrix.}
        \label{fig/sim_only_be-like_042623_atomdb313_vs_spex30801_db1idx_1-100_db2idx_1-100}
    \end{subfigure}

    \caption{The total similarity matrix of Fe \textsc{xxiii} ion between AtomDB v3.1.3 and SPEX v3.08.01. The $x$-axis of the two panels represents the indices in SPEX v3.08.01, and the $y$-axis represents the indices in AtomDB v3.1.3. Panel (a) shows the full total similarity matrix. Panel (b) shows the total similarity matrix of the index range of 1 to 100. }
    \label{fig:total_similarity_matrix}
\end{figure}

After calculating the three similarity components, we combine them into a total similarity matrix. For a level pair $(i,j)$ from two databases, the total similarity is defined as
\begin{equation}
S_{ij} = w_{\rm cfg} S_{\rm cfg}(i,j) + w_{\rm SLJP} S_{\rm SLJP}(i,j) + w_E S_E(i,j),
\end{equation}
where $S_{\rm cfg}$ is the configuration similarity, $S_{\rm SLJP}$ is the similarity based on $2S+1$, $L$, $J$, and parity $\pi$, and $S_E$ is the energy similarity. 
In this work, we adopt the weights
\begin{equation}
(w_{\rm cfg}, w_{\rm SLJP}, w_E)=(0.47,0.25,0.28).
\end{equation}
These weights are based on empirical findings derived from experiments, which have preserved the contributions of all three similarity components very well. Their robustness against variations in the weight ratios is examined in Sect.~\ref{sec:weight_sensitivity}. The resulting matrix $S$ contains the similarity scores for all possible energy-level pairs between the two databases.
An example is shown in Fig.~\ref{fig:total_similarity_matrix}.

However, not every level in one database should have a counterpart in another database. If the similarity matrix is used directly for assignment, the algorithm may force low-similarity levels to be matched. This is not physical, because one database can contain extra energy levels so that an energy level can have no reasonable counterpart. To avoid this problem, we develop a threshold $\tau$ applied to $S$ before the final matching step.

The threshold defines the minimum similarity required for a real energy-level pair to be considered. We define the profit matrix as
\begin{equation}
P_{ij}=S_{ij}-\tau,~0<\tau<1.
\end{equation}

The matrix $P$ or $S$ indicates likely correspondences, but our goal is to obtain a one-to-one matching map filled with 0 or 1. Therefore we formulate the conversion of the profit matrix as a post-processing problem and we use the Hungarian algorithm to solve it \citep{Kuhn1955,Munkres1957}. The Hungarian algorithm can find the globally optimal one-to-one assignment for a given profit matrix. This is suitable for our problem because our profit matrix are moderate in size and we prefer a global optimum than a local optimum.

Let \(S\) be the total similarity matrix between database \(A\) and database \(B\). 
We define a binary assignment matrix \(X\), where
\begin{equation}
X_{ij} =
\begin{cases}
1, & \mathrm{if~level~} i \mathrm{~in~database~} A \mathrm{~is~matched~to~level~} j \mathrm{~in~database~} B,\\
0, & \mathrm{otherwise}.
\end{cases}
\end{equation}
The one-to-one constraint requires
\begin{equation}
\sum_j X_{ij} \leq 1, \qquad
\sum_i X_{ij} \leq 1 .
\end{equation}
Without threshold filtering, the assignment maximizes the total selected similarity,
\begin{equation}
\max_X \sum_{i,j} S_{ij}X_{ij}.
\end{equation}
This gives a global one-to-one assignment, but it can still force weak matches. With a threshold \(\tau\), the profit of selecting a real pair is defined as \(P_{ij}=S_{ij}-\tau\), and the optimization becomes
\begin{equation}
\max_X \sum_{i,j} (P_{ij})X_{ij}=\max_X \sum_{i,j} (S_{ij}-\tau)X_{ij},
\end{equation}
under the same one-to-one constraints. A pair is kept only if it is better than leaving the corresponding levels unmatched. In practice, this is implemented by expanding the matrix with dummy rows and columns that represent unmatched levels. Finally, we only retain the real energy-level pairs.

Thus, $\tau$ serves as a passing threshold. The Hungarian algorithm selects the one-to-one set of pairs that maximizes the total profit relative to $\tau$. In this way, the method finds the best matching map rather than a forced full assignment. The example equation~\ref{eq:threshold_example} illustrates the difference between the assignment before and after applying the threshold $\tau$.
\begingroup
\small
\begin{equation}
\begin{bmatrix}
0.95 & \boxed{0.72} & 0.10\\
\boxed{0.78} & 0.40 & 0.30\\
0.20 & 0.35 & \boxed{0.65}
\end{bmatrix}
\xrightarrow{\;\tau=0.70,\;P=S-\tau\;}
\begin{bmatrix}
0.25 & 0.02 & -0.60\\
0.08 & -0.30 & -0.40\\
-0.50 & -0.35 & -0.05
\end{bmatrix}
\Longrightarrow
\begin{bmatrix}
\boxed{1} & 0 & 0\\
0 & 0 & 0\\
0 & 0 & 0
\end{bmatrix}.\label{eq:threshold_example}
\end{equation}
\endgroup
In this case, the threshold-constrained Hungarian matching keeps only the strongest pair, rather than accepting matches ($S=0.72$ and $0.78$) that only slightly exceed the threshold. The threshold $\tau$ therefore acts as a physical filter on the similarity matrix. 

Repeating this process with different values of \(\tau\) produces different matching maps. A low threshold gives a larger but less conservative map, while a high threshold gives a smaller and more reliable map. This allows the user to choose a suitable threshold for the different use case. The evaluation of the matching maps and the choice of suitable thresholds are discussed in Sect.~4.

\section{Results}

The cross-database energy-level matching map, validation results with different $\tau$, and supporting materials obtained in this work are presented below. The standardized atomic data, trained configuration models, source codes, and interactive tools used to reproduce these results are provided in the CFADB-XCOS archive. A temporary review copy is available through Zenodo Sandbox.\footnote{\url{https://sandbox.zenodo.org/records/643614}} Before final publication, the archive will be deposited in the production Zenodo repository and the temporary link will be replaced by the corresponding
persistent DOI.

\subsection{Rule-Based Validation and Acceptable $\tau$}

The threshold $\tau$ in the Hungarian assignment controls how many energy-level pairs are kept in the final map. A low threshold produces more matches, but it may include weak or unphysical pairs. A high threshold gives a more conservative map, but it may remove real correspondences. We therefore scan $\tau$ and evaluate the matching maps at each $\tau$.

To judge whether a selected energy-level pair is physically consistent, we define six validation rules for a candidate pair \((i,j)\):
\begin{itemize}
    \item[(1)] The configuration similarity must be above a chosen threshold,
    \begin{equation}
    S_{\rm cfg}(i,j) \geq S_{\rm cfg}^{\rm c}.
    \end{equation}

    \item[(2)] The spin multiplicity must be exactly the same,
    \begin{equation}
    (2S+1)_i = (2S+1)_j .
    \end{equation}

    \item[(3)] The total orbital angular momentum quantum number must be exactly the same,
    \begin{equation}
    L_i = L_j .
    \end{equation}

    \item[(4)] The total angular momentum quantum number must be exactly the same,
    \begin{equation}
    J_i = J_j .
    \end{equation}

    \item[(5)] The parity $\pi$ must be valid and exactly the same,
    \begin{equation}
    \pi_i=\pi_j,\qquad \pi_i,\pi_j\in\{0,1,2\}.
    \end{equation}

    \item[(6)] The difference of energy value $E$ must not exceed a chosen threshold. 
    We define the relative difference of $E$ as
    \begin{equation}
    \Delta E_{\rm rel}(i,j)
    =
    \frac{|E_i-E_j|}
    {\max(|E_i|,|E_j|,\epsilon)}
    \leq E_{\rm rel}^{\rm c},
    \end{equation}
    and the absolute difference of $E$ as
    \begin{equation}
    \Delta E_{\rm abs}(i,j)
    =
    |E_i-E_j|
    \leq E_{\rm abs}^{\rm c}.
    \end{equation}
    The energy condition can be applied in four modes: relative-only, absolute-only, relative-or-absolute, and relative-and-absolute, which means that a candidate pair will be retained if it satisfies the relative criterion ($\Delta E_{\rm rel}\leq E_{\rm rel}^{\rm c}$), the absolute criterion ($\Delta E_{\rm abs}\leq E_{\rm abs}^{\rm c}$), either criterion, or both criteria, respectively.
\end{itemize}

In this work, H-like, He-like, Li-like, and Be-like ions use the relative-only energy rule, while Ne-like ions use the relative-or-absolute rule, with the thresholds listed in Table~\ref{tab:seq_validation_parameters}. This is useful because low energy and high energy levels are not always best described by the same energy criterion. As shown in Table~\ref{tab:energy_difference}, the relative criterion is more suitable for the Fe \textsc{xvii} case with \(\Delta E_{\rm rel}=0.014\) and \(\Delta E_{\rm abs}=14.07\), whereas the absolute criterion is more suitable for the Mg \textsc{iii} case with \(\Delta E_{\rm rel}=0.064\) and \(\Delta E_{\rm abs}=3.39\).

\begin{table}[hbt!]
    \centering
    \caption{Differences in $\Delta E_{\rm rel}$ and $\Delta E_{\rm abs}$ for selected matched energy-level pairs in Ne-like ions. For each selected matched level pair, the two levels from different databases have exactly the same configuration and identical $2S+1,~L,~J$, and parity $\pi$. $E_1$ and $E_2$ are the energy values in eV from AtomDB v3.1.3 and SPEX v3.08.01, respectively.}
    \label{tab:energy_difference}
    \begin{tabular}{cccccc}
        \hline
        Ion & Term & $E_1$ (eV) & $E_2$ (eV) & $\Delta E_{\rm rel}$ & $\Delta E_{\rm abs}$ (eV) \\
        \hline
        Fe \textsc{xvii} & 1s$^2$2s$^2$2p$^5$4f $^3$F$_{3}$ & 1014.13 & 1028.20 & 0.014 & 14.07 \\
        Mg \textsc{iii} & 1s$^2$2s$^2$2p$^5$3s $^3$P$_{2}$ & 49.38 & 52.77 & 0.064 & 3.39 \\
        \hline
    \end{tabular}
\end{table}

We define a rule-consistency matrix \(R\), where
\begin{equation}
R_{ij} =
\begin{cases}
1, & \mathrm{if~the~pair~}(i,j)\mathrm{~passes~all~six~validation~rules},\\
0, & \mathrm{otherwise}.
\end{cases}
\end{equation}
This matrix was calculated after performing a complete pairwise comparison of the two databases, without any threshold filtering or Hungarian assignment applied. From this matrix, we compute a rule-based maximum one-to-one overlap between the two databases, which is called the theoretical match count $N_{\rm theo}$. It is defined as the maximum number of non-repeated energy-level pairs that can be selected from candidates that satisfy the validation rules:
\begin{equation}
N_{\rm theo}
=
\max_Y \sum_{i,j} Y_{ij},
\end{equation}
subject to
\begin{equation}
Y_{ij} \leq R_{ij},\qquad
\sum_j Y_{ij}\leq 1,\qquad
\sum_i Y_{ij}\leq 1,\qquad
Y_{ij}\in\{0,1\}.
\end{equation}
Here \(Y\) is a binary assignment matrix. The constraint \(Y_{ij}\leq R_{ij}\) ensures that only pairs passing all six validation rules can contribute to \(N_{\rm theo}\). Consequently, the \(N_{\rm theo}\) is computed from candidate pairs that simultaneously satisfy the configuration, \(2S+1\), \(L\), \(J\), $\pi$, and energy criteria. This number is a rule-based reference value computed independently of the threshold-constrained Hungarian matching map, and is used as an additional criterion to judge whether the final map recovers the largest possible set of rule-validation consistent one-to-one correspondences.

For each threshold \(\tau\), we apply the threshold-constrained Hungarian assignment described in Sect.~3.4 and obtain a selected matching map, which is a binary matrix \(X(\tau)\). We then compute the number of selected matches,
\begin{equation}
N_{\rm match}(\tau)=\sum_{i,j}X_{ij}(\tau),
\end{equation}
and the number of selected pairs that fail the validation rules
\begin{equation}
N_{\rm mismatch}(\tau)
=
\sum_{i,j}X_{ij}(\tau)\,[1-R_{ij}] .
\end{equation}
In this context, we consider a threshold $\tau$ is accepted when
\begin{equation}
N_{\rm theo}>0,\qquad N_{\rm match}(\tau)=N_{\rm theo},\qquad N_{\rm mismatch}(\tau)=0. \label{N_match=M_theo}
\end{equation}
In other words, we have established a set of evaluation criteria to determine whether the matching map obtained by threshold-constrained Hungarian algorithm is acceptable.
If the map contains the expected number of rule-consistent one-to-one pairs and no pair violates the validation rules, we consider this matching map acceptable, which means that the corresponding threshold is set appropriately.
 If a single threshold meets these conditions, we report it as an acceptable threshold $\tau$. If several consecutive thresholds satisfy these conditions, we report them as an acceptable threshold range.

\subsection{Representative Matching Examples}

We now show representative examples of the threshold-selected matching map. 
For each example, we show the total similarity matrix $S$ and the matching map obtained by threshold-constrained Hungarian algorithm. 
The selected pairs are evaluated with the validation rules described in Sect.~4.1. The rule-validation parameters such as $S_{\rm cfg}^{\rm c},~E_{\rm rel}^{\rm c},~E_{\rm abs}^{\rm c}$ in this subsection are the case specific settings used to generate the displayed maps. The rule-validation parameters used for the complete statistical summary are reported in Table~\ref{tab:seq_validation_parameters}. 

\begin{figure}[hbt!]
    \centering

    \begin{subfigure}[t]{0.49\linewidth}
        \centering
        \includegraphics[width=\linewidth]{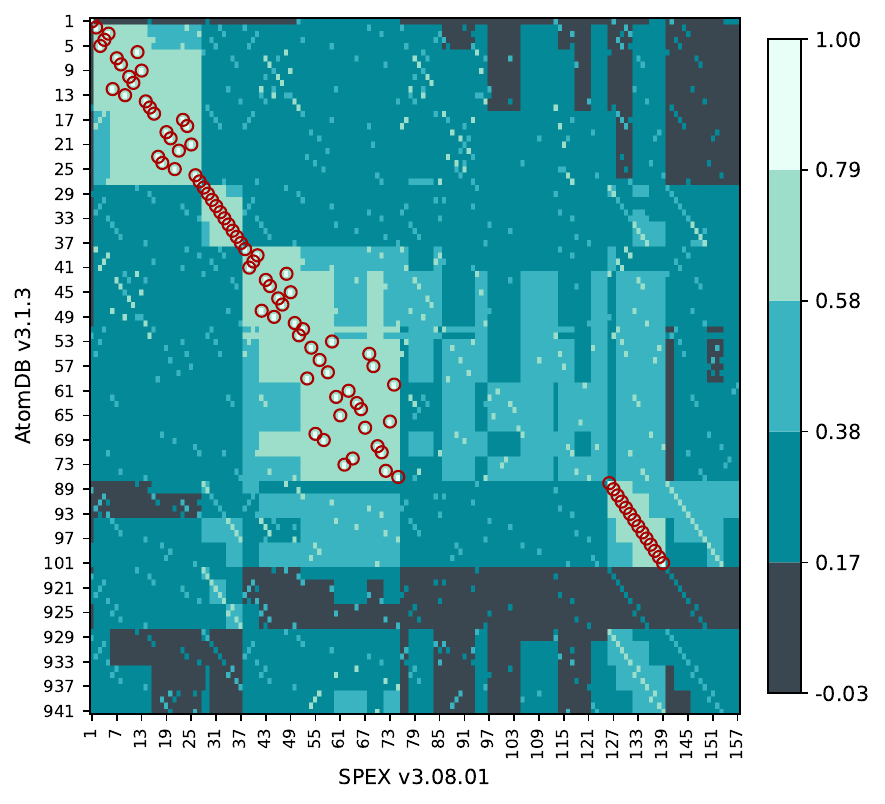}
        \caption{Full similarity matrix and map.}
        \label{fig:102516_atomdb313_vs_spex30801_overlay_thr_0.720000.pdf}
    \end{subfigure}
    \hfill
    \begin{subfigure}[t]{0.49\linewidth}
        \centering
        \includegraphics[width=\linewidth]{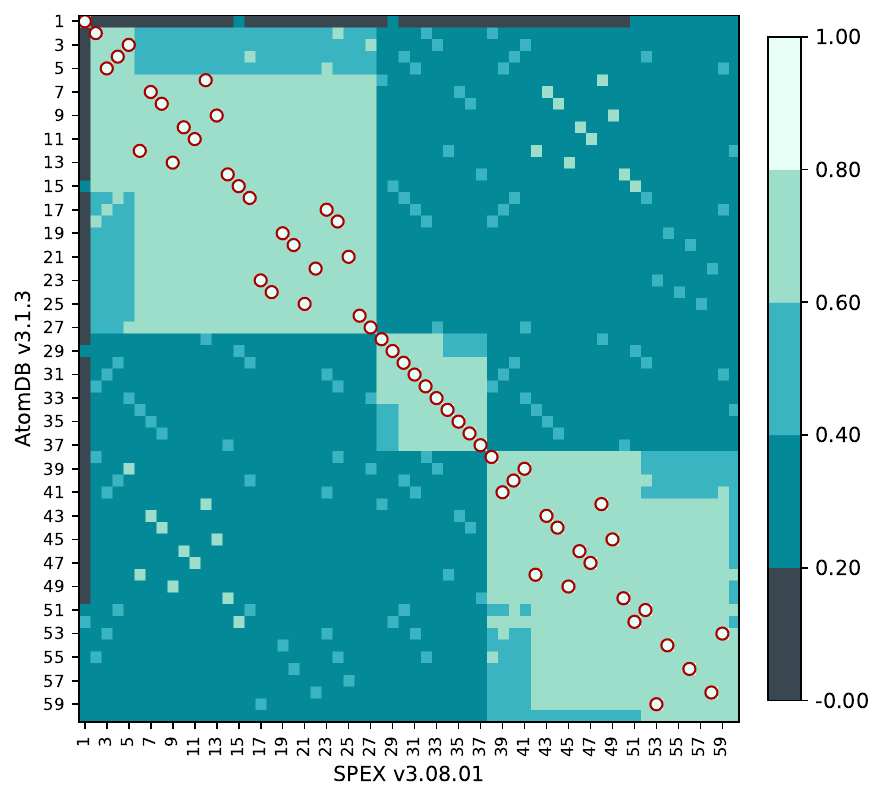}
        \caption{Part of similarity matrix and map.}
        \label{fig/102516_atomdb313_vs_spex30801_overlay_thr_0.720000（1）}
    \end{subfigure}

    \caption{The matching map for Mn \textsc{xvi} between AtomDB v3.1.3 and SPEX v3.08.01 is shown in panels (a) and (b). Panel (a) shows the full similarity matrix, with red-brown ring markers indicating the matching pairs selected by the threshold-constrained Hungarian algorithm. The x-axes and y-axes represent the energy-level indices in SPEX and AtomDB respectively. The rule-validation parameters of this map are exact agreement in $2S+1$, $L$, $J$, and $\pi$, together with \(S_{\rm cfg}\geq0.89\) and \(\Delta E_{\rm rel}\leq0.023\) under the relative-only energy rule. The acceptable threshold range is \(\tau=0.55\text{--}0.98\). Panel (b) shows an enlarged view of panel (a), restricted to indices 1--60 in both databases, so that matching pairs are more clear.}
    \label{fig:example_map}
\end{figure}

As an example, Figure~\ref{fig:example_map} shows the matching map for Mn \textsc{xvi} between AtomDB v3.1.3 and SPEX v3.08.01, obtained using a threshold within the acceptable range. Throughout the acceptable range (\(\tau=0.55-0.98\)), all six validation rules are satisfied, and $N_{\rm match}(\tau)=N_{\rm theo}=89$. 
As mentioned in Sec. 4.1, we consider the matching map obtained within the acceptable threshold range to be the most reasonable map under this criterion. We have identified acceptable threshold ranges for the H-like, He-like, Li-like, Be-like, and Ne-like sequences, which correspond to the most reasonable maps. We will present the specific statistical results in Section 4.3.
 
\begin{figure}[hbt!]
    \centering

    \begin{subfigure}[t]{0.49\linewidth}
        \centering
        \includegraphics[width=\linewidth]{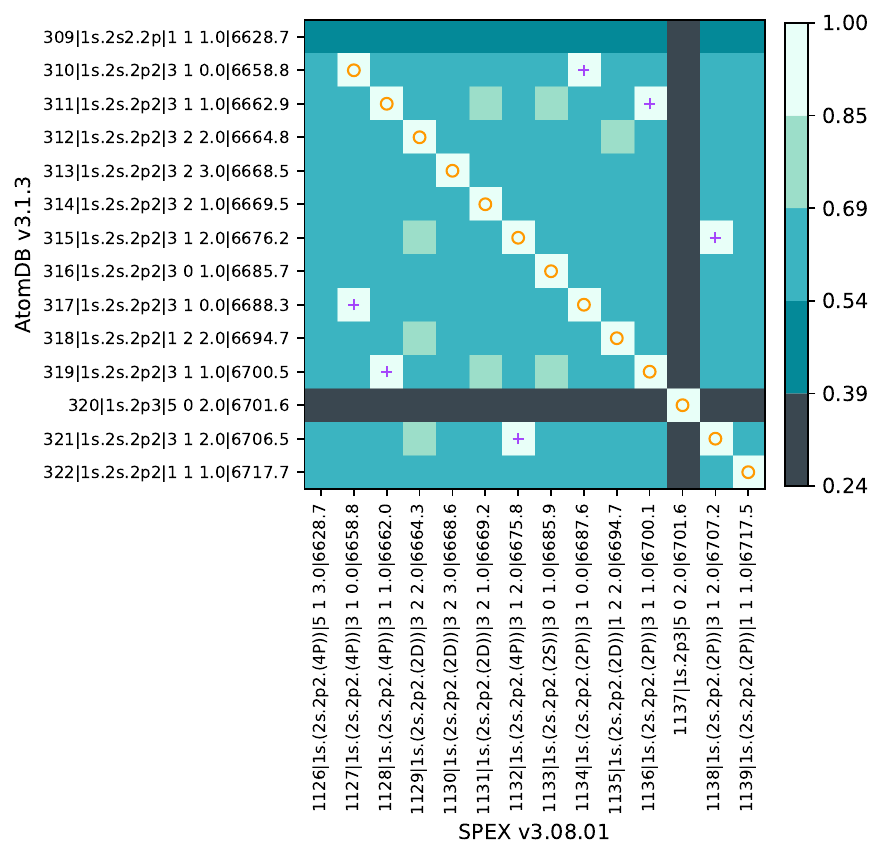}
        \caption{Complex parent-label notation.}
        \label{fig:042623_atomdb313_vs_spex30801_overlay_thr_0.890000_diff_rules}
    \end{subfigure}
    \hfill
    \begin{subfigure}[t]{0.49\linewidth}
        \centering
        \includegraphics[width=\linewidth]{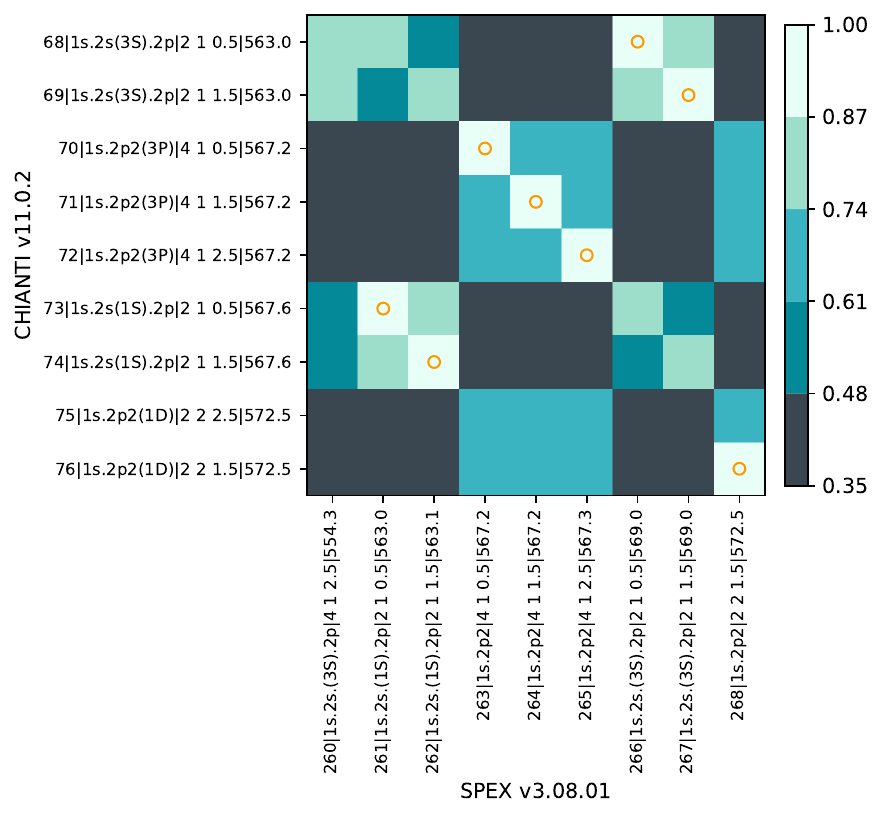}
        \caption{Configuration--energy competition.}
        \label{fig/030806_chianti1102_vs_spex30801_overlay_thr_0.730000_diff_rules}
    \end{subfigure}

    \caption{The matching maps obtained by threshold-constrained Hungarian algorithm for two relatively difficult cross-database cases are shown in panels (a) and (b). Panel (a) shows the matching map for Fe \textsc{xxiii} between AtomDB v3.1.3 and SPEX v3.08.01. The background represents the similarity matrix. The orange ring markers denote matching pairs that satisfy all six validation rules, whereas purple plus markers denote candidate pairs that meet the rules but were not matched. The compact label for example ``309|1s.2s2.2p|1 1 1.0|6628.7'' represents index 309, configuration 1s2s$^2$2p, $2S+1=1$, $L=1$, $J=1$, and $E=6628.7$ eV. The parity $\pi=1$ in this example, is included in the rule-validation although it is not printed separately in labels. The map displayed here employs a context-specific validation setup that requires exact agreement in $2S+1$, $L$, $J$, and $\pi$, together with \(S_{\rm cfg}\geq0.80\) and \(\Delta E_{\rm rel}\leq0.11\) under the relative-only energy rule; its acceptable threshold range is \(\tau=0.81\text{--}0.90\). Panel (b) shows the matching map for O \textsc{vi} between CHIANTI v11.0.2 and SPEX v3.08.01; compared with panel (a), it uses \(S_{\rm cfg}\geq0.86\), \(\Delta E_{\rm rel}\leq0.12\), and an acceptable threshold range of \(\tau=0.73\text{--}0.93\), while the other validation and plotting conventions remain unchanged.}
    \label{fig:difficult_matching_examples}
\end{figure}

We further illustrate how acceptable thresholds can produce physically consistent matching maps under complex conditions.
Figure~\ref{fig:difficult_matching_examples} shows two examples obtained under relatively difficult matching conditions. Panel (a) shows the matching map for Fe \textsc{xxiii} between AtomDB v3.1.3 and SPEX v3.08.01 in the acceptable threshold range. This case illustrates a situation where one database contains more complex configurations, including multiple parentheses and parent-state labels. Even in this case, the accepted map selects pairs with consistent configuration, \(2S+1\), \(L\), \(J\), $\pi$, and similar energy values. This shows that the configuration model and the threshold-constrained Hungarian assignment can match configurations that are expressed differently but are physically identical. Furthermore, we can see that the energy values corresponding to the selected pair show the highest match among the relevant candidate solutions. This once again demonstrates that our matching method is reasonable and effective.

Figure~\ref{fig:difficult_matching_examples}(b) shows the matching map for O \textsc{vi} between CHIANTI v11.0.2 and SPEX v3.08.01 in the acceptable threshold range. This case focuses on a local region of the map where the configuration and the energy values are not fully aligned. Some candidate pairs have matching configurations and parent-state labels, but their energies are not the closest possible ones. 
For example, pairs such as \(73 \leftrightarrow 261\), \(74 \leftrightarrow 262\), \(68 \leftrightarrow 266\), and \(69 \leftrightarrow 267\) keep the same parent-state labels, and $S,~L,~J,~\pi$. However, if we consider the energy values, the difference between the other pairs such as \(73 \leftrightarrow 266\), \(74 \leftrightarrow 267\), \(68 \leftrightarrow 261\), and \(69 \leftrightarrow 262\) would be even smaller. 
Although the difference in energy values is relatively large compared to the other pairs, the selected pairs still satisfy the adopted energy criterion \(\Delta E_{\rm rel}\leq0.12\), so they are accepted by the validation rules. If a user prefers a more energy-dominated map, these cases can be quickly identified by tightening \(\Delta E_{\rm rel}\) or requiring an absolute energy threshold \(\Delta E_{\rm abs}\) to be satisfied as well, and then modifying the released map if needed.

\begin{figure}[hbt!]
    \centering
    \includegraphics[width=0.7\linewidth]{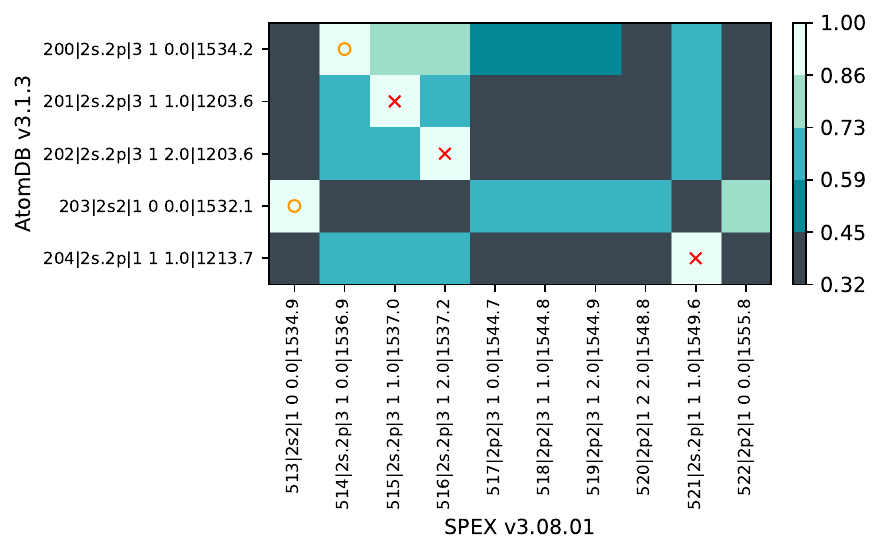}
    \caption{The matching map for F \textsc{viii} between AtomDB v3.1.3 and SPEX v3.08.01 at \(\tau=0.90\), below the acceptable range of \(\tau=0.96\text{--}0.99\). Orange circles indicate selected Hungarian matches that meet all six validation rules, while red crosses indicate candidate pairs that do not meet the rules but were selected via one-to-one mapping. “200|2s.2p|3 1 0.0|1534.2” means the index is 200, the configuration is 2s2p, $2S+1=3,~L=1,~J=0$, and $E=1534.2$ eV. The energy values of the matches marked with red crosses are highly inconsistent. When the threshold is set within 0.96--0.99, all matches marked with red crosses are deselected.}
    \label{fig:020908_atomdb313_vs_spex30801}
\end{figure}

Fig.~\ref{fig:020908_atomdb313_vs_spex30801} shows the matching map for F \textsc{viii} between AtomDB v3.1.3 and SPEX v3.08.01 below the acceptable threshold range. Among the three matching pairs \(201 \leftrightarrow 515, 202 \leftrightarrow 516, \text{and } 204 \leftrightarrow 521\), the configurations, \(2S+1\), \(L\), \(J\), and $\pi$ are identical, but their energy values differ by approximately 336 eV (about 21.7\%). However, due to the high similarity in other parts, their overall similarities reach 0.934, 0.934, and 0.959, respectively. When we increase the threshold to the acceptable range of 0.96--0.99, these three matches are no longer selected. This demonstrates that our method can distinguish and exclude pairs where the configuration, \(2S+1\), \(L\), \(J\), and $\pi$ are identical but the energy values differ significantly. At the same time, if users believe these matches should actually be selected, they can choose results that fall below the acceptance threshold, which shows that our method also gives users great flexibility.

\subsection{Summary of Parameters and $\tau$ in 5 Sequences}

We applied the threshold-constrained Hungarian algorithm  to H-like, He-like, Li-like, Be-like and Ne-like. A database pair was included whenever both databases contain energy-level data for the ion. This gives a total of 314 database pairs in all ions: 71 in H-like, 65 in He-like, 64 in Li-like, 64 in Be-like, and 50 in Ne-like. 
First, we determined the rule-validation parameters for all these sequences. They are summarized in Table~\ref{tab:seq_validation_parameters}. These parameters are the parameters of the six validation rules mentioned in Section 4.1. 
The value of \(S_{\rm cfg}^c\) is set to 0.01 below the minimum cosine similarity among the hard, shortening, and medium pairs for each sequence, where \(S_{\rm cfg}^c = \min\{S_{\rm hard},S_{\rm shortening},S_{\rm medium}\}-0.01\). Figure~\ref{fig:cfg_cosine_boxplot} shows these distributions for Li-like ions. Because the H-like and He-like training sets contain no shortening or medium pairs, their $S_{\rm cfg}^c$ are defined as \(S_{\rm cfg}^c = S_{\rm hard}-0.01\). 
Since the hard, shortening, and medium cases represent configurations that can be matched physically, we require level pairs satisfying these conditions to be included among the candidate pairs. \(\Delta E_{\rm rel}^c\) and \(\Delta E_{\rm abs}^c\) are the minimum values obtained by scanning \(\Delta E_{\rm rel}^c\) and \(\Delta E_{\rm abs}^c\) when there is an acceptable threshold range. Under the most stringent criteria, we still identified an acceptable threshold range for every available database pair across all ions, indicating that our method produces reasonable matching maps in all cases.

\begin{table}[h]
\centering
\caption{The parameters of the six validation rules used for all isoelectronic sequences. A value of 1 in the \(2S+1,~L,~J,~\pi\) column means that they should be exactly identical for the matching pair. The reserved value \(\pi=2\) is used only for matching two \texttt{limit} entries, while \(\pi=-1\) does not pass validation.}
\label{tab:seq_validation_parameters}
\begin{tabular}{lccccc}
\hline
Sequence & \(S_{\rm cfg}^c\) & \(2S+1,~L,~J,~\pi\) & \(\Delta E_{\rm rel}^c\) & \(\Delta E_{\rm abs}^c\) (eV) & Energy rule mode\\
\hline
H-like   & 0.99 & 1 & 0.005 & -   & only \(\Delta E_{\rm rel}\)\\
He-like  & 0.99 & 1 & 0.008 & -   & only \(\Delta E_{\rm rel}\)\\
Li-like  & 0.85 & 1 & 0.013 & -   & only \(\Delta E_{\rm rel}\)\\
Be-like  & 0.80 & 1 & 0.064 & - & only \(\Delta E_{\rm rel}\)\\
Ne-like  & 0.76 & 1 & 0.026 & 4.3 & \(\Delta E_{\rm rel}\) or \(\Delta E_{\rm abs}\)\\
\hline
\end{tabular}
\end{table}

\begin{figure}[hbt!]
    \centering
    \includegraphics[width=0.95\linewidth]{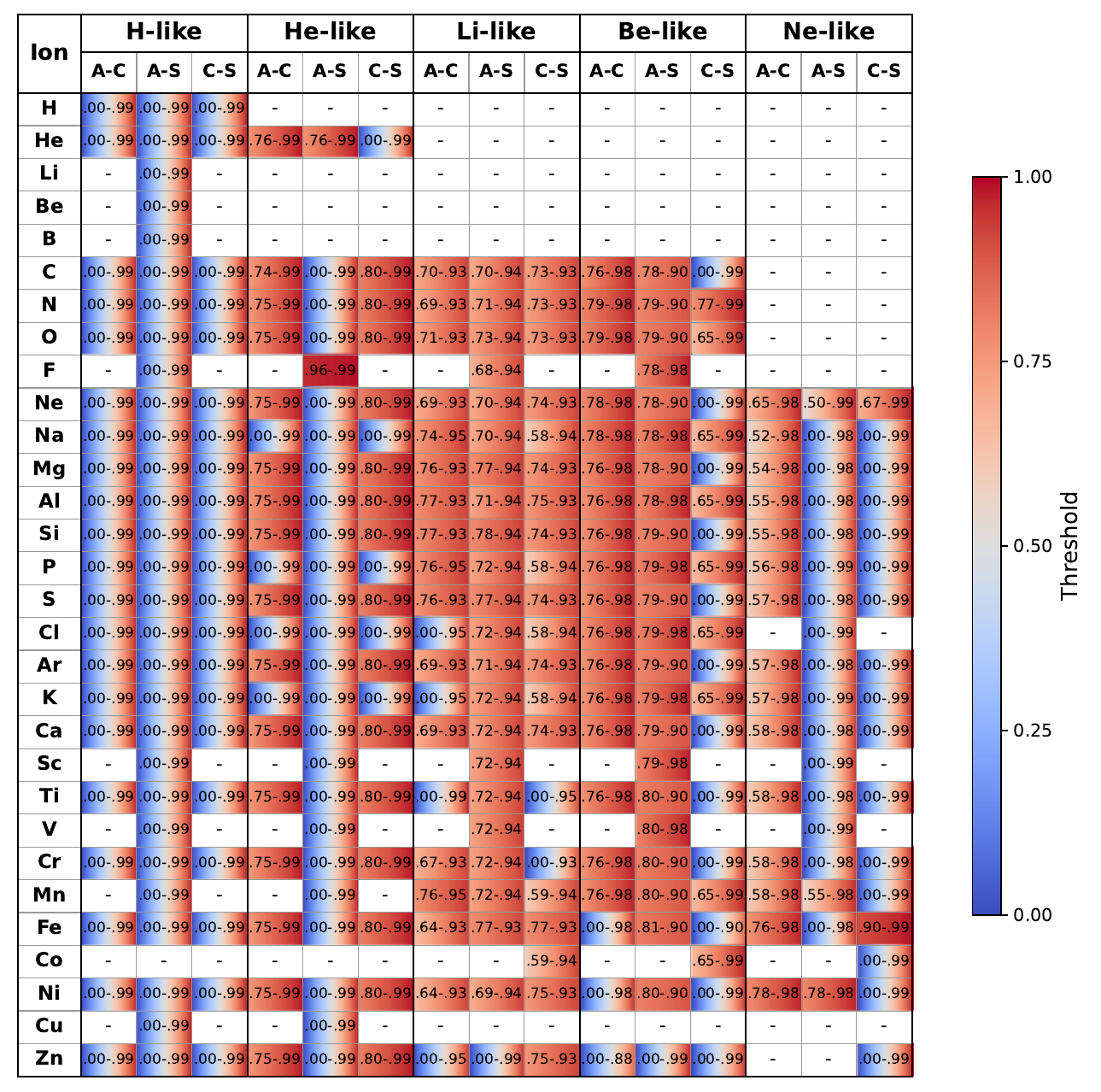}
    \caption{Statistics of the acceptable threshold ranges of all ions. The sequence-dependent parameters are shown in Table~\ref{tab:seq_validation_parameters}. The “A,” “C,” and “S” in the second row represent AtomDB v3.1.3, CHIANTI v11.0.2, and SPEX v3.08.01, respectively. The colour gradient represents the threshold values between these endpoints. The dash indicates that there is insufficient data to match, not that the match failed.}
    \label{fig:accepted_threshold_ranges}
\end{figure}

Then we scanned the threshold from 0 to 1 for all ions under these six validation rules, including exact standardized parity-label consistency. 
The acceptable threshold ranges are summarized in Fig.~\ref{fig:accepted_threshold_ranges} for every ion using the rule-validation parameters shown in Table~\ref{tab:seq_validation_parameters}. 
This acceptable threshold range may indicate the stability of the map with respect to threshold selection. A broad range indicates that the matching result is stable, while a narrow range indicates that the map is more sensitive to the threshold choice. This range is also provided as a reference here to help researchers quickly select appropriate thresholds.

As shown in Fig.~\ref{fig:accepted_threshold_ranges}, the acceptable threshold ranges are generally broader for H-like ions. This is likely because H-like ions have fewer energy-level data, simpler atomic structure, and clearer one-to-one correspondences among different databases. 
This also suggests that different isoelectronic sequences do not necessarily require the same rule-validation parameters. The acceptable threshold ranges vary with the isoelectronic sequence, database pair, and ion. These differences may be related to several factors, including the coverage and number of energy-level data in each database, the labels of configurations, and the complexity of the atomic structure.

We also repeated the threshold scan after replacing SPEX v3.08.01 with SPEX v3.08.03 (Kaastra et al. 2026) with the rule-validation parameters in Table~\ref{tab:seq_validation_parameters}. All matching maps again have nonempty acceptable threshold ranges. 
The number of matching pairs and acceptable ranges are unchanged for H-like and He-like ions. However, the SPEX-involving matching map have a larger number of pairs and different acceptable threshold ranges for Li-like, Be-like, and Ne-like ions. 
Thus, there are still acceptable matching maps in this version comparison, while the sizes of the maps and their acceptable threshold ranges can depend on the database version.

\subsection{Weight Sensitivity and Component Ablation}
\label{sec:weight_sensitivity}

We examined the robustness of the total similarity weights while keeping the rule-validation parameters in Table~\ref{tab:seq_validation_parameters} fixed. We scan the non-negative weights over 
\begin{equation}
w_{\rm cfg}+w_{\rm SLJP}+w_E=1
\end{equation}
with a step size of 0.1, giving 66 grid points, and the adopted point$(0.47,0.25,0.28)$ was evaluated separately. The analysis tests weight robustness rather than identifying a unique optimum.

Here, we identify one case as the matching of one ion between one available database pair, such as O~\textsc{vi} between AtomDB v3.1.3 and SPEX v3.08.01. The five isoelectronic sequences considered here contain 314 such cases. For a given weight triplet, a case is successful if at least one  threshold satisfies the condition in Eq.~\ref{N_match=M_theo}. 
The quantities $N_{\rm theo}$ and $N_{\rm mismatch}$ are evaluated using the six validation rules. Cases with $N_{\rm theo}=0$ are not counted as successful. 

\begin{figure}[hbt!]
    \centering
    \includegraphics[width=0.8\linewidth]{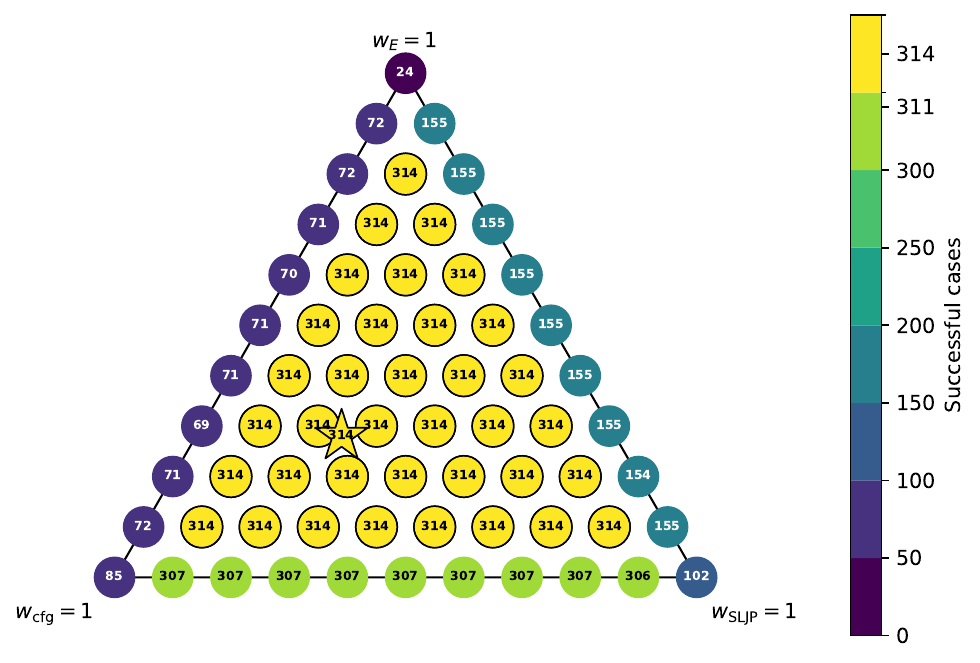}
    \caption{
    Weight sensitivity of the total similarity score in ternary weight space. Each circular point represents one of the 66 weight triplets $(w_{\rm cfg},w_{\rm SLJP},w_E)$ sampled with spacing 0.1 and satisfying $w_{\rm cfg}+w_{\rm SLJP}+w_E=1$. The three components describe configuration,  quantum numbers and parity $(2S+1,L,J,\pi)$, and energy values, respectively. The top, bottom-left, and bottom-right vertices correspond to energy-only, configuration-only, and SLJP-only scores.  Points in non-vertex edge represent two-component scores. Colours and numerical labels give the number of successful cases out of 314. The star marks the adopted weights $(0.47,0.25,0.28)$ in our method, which succeed in all 314 cases.}
    \label{fig:weight_sensitivity_success}
\end{figure}

Figure~\ref{fig:weight_sensitivity_success} shows that the adopted weights succeed in all 314 cases. All 36 sampled interior points, defined by $w_{\rm cfg}>0$, $w_{\rm SLJP}>0$, and $w_E>0$, also achieve complete coverage, whereas none of the 30 sampled boundary points does. 
Therefore, complete coverage proves to be highly robust across all tested combinations of internal weights. Furthermore, complete coverage can only be achieved when all three components contribute.

The configuration-only, SLJP-only, and energy-only scores succeed in 85, 102, and 24 cases, respectively. At the equal-weight two-component points $(0.5,0.5,0)$, $(0.5,0,0.5)$, and $(0,0.5,0.5)$, the configuration-plus-SLJP, configuration-plus-energy, and SLJP-plus-energy variants succeed in 307, 71, and 155 cases, respectively, whereas the adopted three-component score succeeds in all 314 cases. These results show that configuration and $S,~L,~J,~P$ together provide the information needed for most cases, while the energy component supplies complementary information required to achieve complete coverage under the tested settings. Omitting either configuration or $S,~L,~J,~P$ substantially reduces the number of successful cases.

\FloatBarrier
\section{Summary}

In this work, we developed an automated framework for cross-database mapping of atomic energy-level data. The purpose is to identify corresponding energy levels between different atomic databases, which is a necessary first step for comparing transition data, atomic process data, and ultimately plasma emission models.

We applied the framework to AtomDB v3.1.3, CHIANTI v11.0.2, and SPEX v3.08.01. The original data were converted into a standardized format containing the index, configuration, \(2S+1\), \(L\), \(J\), parity $\pi$, and energy values. The matching method combines three complementary similarity components: configuration similarity calculated by a BiLSTM-based model, a rule-based similarity for \(2S+1\), \(L\), \(J\), and parity $\pi$, and an energy values similarity based on the transformed absolute energy, the energy group, and the relative position within that group. These components are combined into a total similarity matrix, and a threshold-constrained Hungarian algorithm is used to obtain a globally optimal partial one-to-one matching map.

We introduced a rule-based validation procedure to evaluate the matching maps. The six validation rules check the configuration similarity, exact consistency in \(2S+1\), \(L\), \(J\), and $\pi$, and the adopted energy-difference criterion. The validation provides a reference value to select an acceptable threshold to obtain an acceptable matching map. We tested the method on H-like, He-like, Li-like, Be-like, and Ne-like isoelectronic sequences. For the ions and database pairs considered in this work, acceptable threshold ranges were found under the adopted rule-validation parameters. Representative examples show that our method can handle complex configuration notation and parent-state labels, and can also identify cases where configuration and energy values give slightly different preferences.

This work does not aim to determine which database is more accurate. Instead, it provides a reusable method to match the cross-database energy-level data. The matching maps can serve as a basis for future comparisons of wavelengths, transition probabilities, collision strengths, and the differences between astrophysical plasma emission models. Future work will extend this framework to other more complex isoelectronic sequences, where energy-level data matching is expected to be more difficult because of denser structures and more complicated configuration labels. For these cases, we will explore machine learning based mapping methods that use both energy-level data and transition data to determine the final cross-database matching maps (Long et al., in prep.).

\begin{acknowledgements}
Junjie Mao acknowledges support from the Young Data Scientist Project of the National Astronomical Data Center (NADC; Grant No. NADC2024YDS-01). Dongwei Fan acknowledges support from the National Natural Science Foundation of China (NSFC; Grant No. 12273077).
CHIANTI is a collaborative project involving George Mason University, the University of Michigan (USA), the University of Cambridge (UK), and NASA Goddard Space Flight Center (USA).
AI tools were used only to assist with code development, and code was reviewed by Qirui Long.
\end{acknowledgements}

\bibliography{bibtex}
\bibliographystyle{raa}

\label{lastpage}

\end{document}